\PassOptionsToPackage{dvipsnames}{xcolor}
\documentclass[sigconf]{acmart}
\usepackage{adjustbox}
\usepackage{algorithm}
\usepackage{algorithmic}
\usepackage{amsmath,amsfonts}

\usepackage{amssymb}
\usepackage{array}
\usepackage{booktabs}
\usepackage[skip=1pt,labelfont=bf]{caption}
\usepackage{calc}
\usepackage{calligra}
\usepackage{color}
\usepackage{colortbl}
\usepackage{courier}
\usepackage{csvsimple}
\usepackage{enumitem}
\usepackage{fancybox}
\usepackage{fontenc}
\usepackage{graphicx}
\usepackage{listings}
\usepackage[dvipsnames]{xcolor}
\usepackage{longtable}
\usepackage{lscape}
\usepackage{makecell}
\usepackage{marvosym}
\usepackage{moreverb}
\usepackage{multicol}
\usepackage{multirow}
\usepackage{pifont}% http://ctan.org/pkg/pifont%\usepackage{subfigure}
\usepackage{pgfplots}
\usepackage{pgf-pie}
\usepackage{rotating}
\usepackage{siunitx}
\usepackage{soul}
\usepackage{subcaption}
\usepackage{tablefootnote}
\usepackage[most]{tcolorbox}
\usepackage{microtype}
\usepackage{threeparttable}
\usepackage{tikz}
\usepackage{url}
\usepackage{wasysym}
\usepackage{xspace}

\usepgfplotslibrary{statistics}
\usetikzlibrary{matrix, shapes.geometric, arrows}
\usetikzlibrary{shapes, arrows, positioning}

\newcolumntype{L}[1]{>{\raggedright\let\newline\\\arraybackslash\hspace{0pt}}m{#1}}
\newcolumntype{C}[1]{>{\centering\let\newline\\\arraybackslash\hspace{0pt}}m{#1}}
\newcolumntype{R}[1]{>{\raggedleft\let\newline\\\arraybackslash\hspace{0pt}}m{#1}}

\definecolor{codegreen}{rgb}{0,0.6,0}
\definecolor{codered}{rgb}{1,0,0}
\definecolor{codegray}{rgb}{0.5,0.5,0.5}
\definecolor{codepurple}{rgb}{0.58,0,0.82}
\definecolor{backcolour}{rgb}{0.95,0.95,0.92}
\definecolor{lightgray}{gray}{0.9}

\newboolean{showcomments}
\setboolean{showcomments}{true}
\ifthenelse{\boolean{showcomments}}
 { \newcommand{\mynote}[2]{
      \fbox{\bfseries\sffamily\scriptsize#1}
        {\small$\blacktriangleright$\textsf{\emph{#2}}$\blacktriangleleft$}}}
        { \newcommand{\mynote}[2]{}}
        
\definecolor{DarkOrange}{rgb}{0.8,0.3,0.0}
\definecolor{DarkCyan}{rgb}{0.0, 0.55, 0.55}
\definecolor{DarkCyel}{rgb}{1.0, 0.49, 0.0}
\definecolor{yellow-green}{rgb}{0.6, 0.8, 0.2}

\newcommand{\etal}{\emph{et~al.}\xspace}
\newcommand{\find}[1]{
\begin{tcolorbox}[leftrule=1mm,toprule=0mm,bottomrule=0mm,left=1pt,right=2pt,top=2pt,bottom=2pt]%[tile,size=fbox,boxsep=2mm,boxrule=0pt,top=0pt,bottom=0pt,borderline={0.5mm}{0pt}{black!70!white},colback=black!5!white]
\em #1
\end{tcolorbox}
}

\lstdefinelanguage{mymarkdown}{
    morekeywords={*,\#, \#\#, \#\#\#},
    sensitive=false,
    morecomment=[l]{//},
    morestring=[b]",
    commentstyle=\color{codegreen},
    keywordstyle=\color{magenta},
    numberstyle=\tiny\color{codegray},
    stringstyle=\color{codepurple},
    basicstyle=\small,
    breakatwhitespace=false,         
    breaklines=true,
    breakindent=0pt,
    keepspaces=true,                 
    numbers=left,                    
    numbersep=5pt,                  
    showspaces=false,                
    showstringspaces=false,
    showtabs=false,                  
    tabsize=2,
}

\lstdefinestyle{mystyle}{
    commentstyle=\color{codegreen},
    keywordstyle=\color{magenta},
    numberstyle=\small\color{black},
    stringstyle=\color{codepurple},
    basicstyle=\scriptsize\ttfamily,
    breakatwhitespace=false,
    breaklines=true,
    captionpos=b,
    keepspaces=true,
    showspaces=false,
    showstringspaces=false,
    showtabs=false,
    tabsize=2
}

\lstdefinelanguage{diff}{
  morecomment=[f][\color{blue}]{@@},     % group identifier
  morecomment=[f][\color{red}]-,         % deleted lines
  morecomment=[f][\color{codegreen}]+,       % added lines
  morecomment=[f][\color{red}]{---}, % Diff header lines (must appear after +,-)
  morecomment=[f][\color{codegreen}]{+++},
  numberstyle=\tiny\color{codegray},
  numbers=left,                    
  numbersep=5pt,         
}

\setlist{noitemsep} %to leave space around whole list

\definecolor{darkpastelred}{rgb}{0.76, 0.23, 0.13}
\definecolor{ao(english)}{rgb}{0.0, 0.5, 0.0}

\definecolor{darkpastelred}{rgb}{0.76, 0.23, 0.13}
\definecolor{ao(english)}{rgb}{0.0, 0.5, 0.0}

\newboolean{useblue}
\setboolean{useblue}{false} % 设置为true来启用蓝色，false为正常颜色

\newcommand{\maybeblue}[1]{%
    \ifthenelse{\boolean{useblue}}%
    {\textcolor{blue}{#1}}%
    {#1}%
}
\setcopyright{none}
\renewcommand\footnotetextcopyrightpermission[1]{}

\setboolean{showcomments}{true}
\ifthenelse{\boolean{showcomments}}
 { \renewcommand{\mynote}[2]{
      \fbox{\bfseries\sffamily\scriptsize#1}
        {\small$\blacktriangleright$\textsf{\emph{#2}}$\blacktriangleleft$}}}
        { 
            \renewcommand{\mynote}[2]{}
        }
\definecolor{DarkOrange}{rgb}{0.8,0.3,0.0}
\definecolor{DarkCyan}{rgb}{0.0, 0.55, 0.55}
\definecolor{DarkCyel}{rgb}{1.0, 0.49, 0.0}
\definecolor{yellow-green}{rgb}{0.6, 0.8, 0.2}

\newcommand{\tool}{\textsc{ReduceFix}\xspace}
\newcommand{\dataset}{\textsc{LFTBench}\xspace}
\newcommand{\datasetpy}{\textsc{LFTBench-Py}\xspace}
\AtBeginDocument{%
  }
\renewcommand{\shortauthors}{Boyang et al.}

\author{Boyang Yang}
\authornote{Co‑first authors who contributed equally to this work.}
\affiliation{%
  \institution{School of Artificial Intelligence(School of Software), Yanshan University}
  \country{China}
}
\email{yby@ieee.org}

\author{Luyao Ren}
\authornotemark[1]
\affiliation{%
  \institution{School of Computer Science, Peking University}
  \country{China}
}
\email{rly@pku.edu.cn}

\author{Xin Yin}
\affiliation{%
  \institution{The State Key Laboratory of Blockchain and Data Security, Zhejiang University}
  \country{China}
}
\email{xyin@zju.edu.cn}

\author{Jiadong Ren}
\affiliation{%
  \institution{School of Artificial Intelligence(School of Software), Yanshan University}
  \country{China}
}
\email{jdren@ysu.edu.cn}

\author{Haoye Tian}
\authornote{Corresponding author.}
\affiliation{%
  \institution{Department of Computer Science, Aalto University}
  \country{Finland}
}
\email{tianhaoyemail@gmail.com}

\author{Shunfu Jin}
\affiliation{%
  \institution{School of Artificial Intelligence(School of Software), Yanshan University}
  \country{China}
}
\email{jsf@ysu.edu.cn}

\acmConference[ICSE'26]{Proceedings of the 48th International Conference on Software Engineering}
{April 2026}{Rio de Janeiro, Brazil}
\acmYear{2026}
\copyrightyear{2026}
\begin{document}
\title{Input Reduction Enhanced LLM-based Program Repair}
\begin{abstract}
Large Language Models (LLMs) have shown great potential in Automated Program Repair (APR).
Test inputs, being crucial for reasoning the root cause of failures, are always included in the prompt for LLM-based APR.
Unfortunately, LLMs struggle to retain key information in long prompts. When the test inputs are extensive in the prompt, this may trigger the “lost-in-the-middle” issue, compromising repair performance.
% \tian{logic gap. Why do they need failure-inducing tests?}\tian{only failure-indcuing tests? how about other tests or other augmented information?}\yang{I've revised the above sentences.}
%To address this, we propose \tool, an end-to-end LLM-based APR approach, incorporating automated input reduction while keeping failure-inducing semantics. 
To address this, we propose \tool, an LLM-based APR approach with a built-in component that automatically reduces test inputs while retaining their failure-inducing behavior.
% \tian{input reduction is not our ultimate goal.}
\tool prompts an LLM to generate a reducer that minimizes failure-inducing test inputs without human effort, and then feeds the reduced failure-inducing inputs to guide patch generation. 

For targeted evaluation, we constructed \dataset, the first long-input APR benchmark with 200 real bugs from 20 programming tasks, each paired with a failure-inducing input whose median size is 1 MB. On this benchmark, \tool shrinks inputs by 89.1\% on average and improves overall pass@10 by up to 53.8\% relative to a prompt that includes the original test, and by 17.6\% compared with omitting the test entirely. Adding the same reduction step to ChatRepair and CREF increases their fix rate by 21.3\% and 2.6\%, respectively, without other changes.
Our gains hold against an ddmin-only reducing template baseline and transfer to repository-level OSS-Fuzz cases.
Ablation studies further highlight the impact of input length and compressed failure information on repair success. 
These results underscore that automatically reducing failing inputs is a practical and powerful complement to LLM-based APR, significantly improving its scalability and effectiveness.

%\tian{I feel the motivation is weak. Problem priority. We should first highlight the importance of why and how long tests impact the LLM-based repair (the problem)}
\end{abstract}

\maketitle
\begin{figure*}[t]
  \centering
  \resizebox{.9\linewidth}{!}{
    \includegraphics[width=\linewidth]{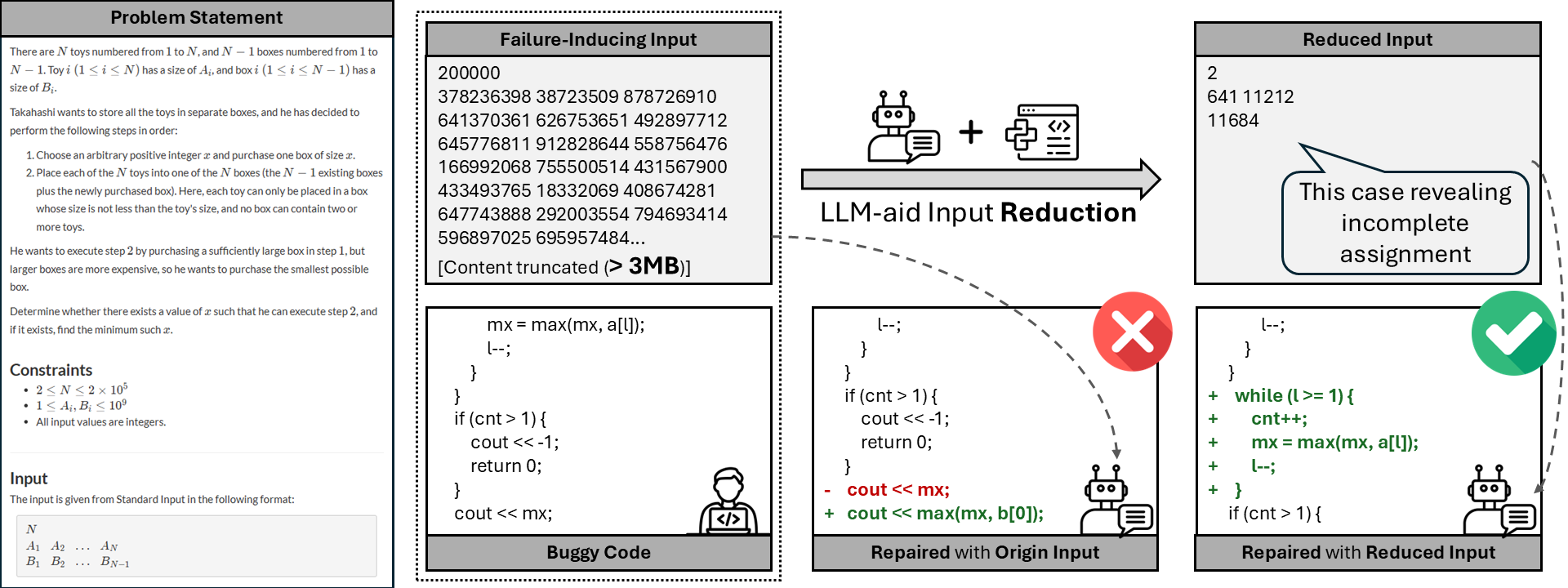}
  }
  \caption{Motivating example (ABC376C): input reduction shrinks a \textbf{>3MB} failure-inducing test to three critical lines, enabling the LLM to generate the accurate patch.}
  \label{fig:example}
\end{figure*}

\section{Introduction}

% reducer is neccessary
APR aims to automatically generate bug-fixing patches for software defects, thereby reducing the manual effort required for debugging~\cite{yang2025surveyllmbasedautomatedprogram,le2019automated,xu2024aligning,chen2025prometheus}. Recently, LLMs have been applied to APR with promising results~\cite{xia2022less,luo2025unlocking}.
Many recent APR systems enhance patch generation by including test inputs in the prompt, among which the test suite plays a particularly important role~\cite{kong2024contrastrepair,tang2024code,yang2024cref}. 
It provides a concrete example of how the program succeeds or fails, helping the LLM focus on the underlying issue and produce a correct fix. 
This strategy has shown strong results in recent systems such as ChatRepair~\cite{xia2024automated}.
% \tian{something like: test input can help LLM to fix bug, because XXX. failure-inducing test is one of them, has been showing good results because of XXX.}\yang{I've revised the first paragraph of introduction.} 
Existing APR studies typically evaluate on benchmarks such as Defects4J~\cite{just2014defects4j} and HumanEval-Java~\cite{jiang_impact_2023}, where test inputs are short and rarely exceed a few hundred characters.
However, when the test input becomes too long, it is difficult to pinpoint the root cause of the error, known as the ``lost-in-the-middle'' phenomenon~\cite{liu2023lost,pham2015hercules}, where information buried in a long prompt receives little attention and overall task performance drops~\cite{yang2024cref,tian2023chatgpt}.
Therefore, an automatic test input reduction step becomes essential before repair.

% labor cost
Existing works on test input reduction rely heavily on human effort, which can be divided into two main categories. 

Syntax-based approaches, such as HDD~\cite{HDD} and Perses~\cite{Perses}, are built on the classical \textit{ddmin} algorithm~\cite{DD} but incorporate grammar or tree structure to ensure syntactic validity during reduction.
However, adapting these APR tools to new tasks requires designing a new grammar and tuning heuristics for each input format, which is both time-consuming and error-prone.
Other techniques, such as ddSMT~\cite{niemetz2013ddsmt} and J-Reduce~\cite{DBLP:conf/pldi/KalhaugeP21}, target specific domains, but still require domain knowledge and significant manual effort to implement and are not reusable across different formats.
In LLM-based APR, the input format varies widely across tasks, often involving different formats ranging from plain text to structured JSON or domain-specific encoding~\cite{liu2020efficiency}.
As a result, relying on hand-crafted reducers is not scalable and makes automation difficult.
 
These observations uncover \emph{two major limitations}:
\begin{itemize}[leftmargin=*]
\item[\ding{172}] \textbf{Lack of length-aware handling for test input.} Although many APR systems embed the full test input into the prompt, few consider how input length affects patch quality. 
CREF~\cite{yang2024cref} finds that on Bard, prompting with the failing test can hurt repair success, achieving 10\% lower accuracy than the no-test baseline in some cases, primarily because long inputs overwhelm the LLM and trigger the ``lost-in-the-middle'' effect~\cite{liu2023lost}.    
However, existing APR systems treat the test input as a fixed block of context and have not attempted to shorten or distill it before repair.

\item[\ding{173}] \textbf{Limitation of manually crafted input reducers.} Prior approaches on input reduction often relies on handwritten syntax grammars or domain-specific rules, which require significant human effort and deep domain knowledge~\cite{HDD,Perses,niemetz2013ddsmt}.  
Even after laborious manual effort, each reducer remains tightly coupled to a specific task or file structure and cannot be generalized across formats.  
As LLM-based APR must handle a wide range of input types, including plain text, JSON, and custom encoding, manual reducers do not scale, and no existing method supports automatic reducer generation across diverse tasks.
\end{itemize}

To tackle the above limitations, we present \tool, an LLM-based program repair framework that integrates automated input reduction into the repair loop.  
The framework prompts the LLM to customize a task-specific reducer, and then applies this reducer to produce a reduced failure-inducing input, finally leverages the reduced input to guide LLMs in generating the correct patch.  
\tool thereby mitigates the ``lost-in-the-middle'' effect by reduction on each test case, allowing the repair model to concentrate on the actual failure-related parts instead of unrelated context.
The overall process follows a three-stage pipeline: (i) reducer generation via a one-shot prompt with the problem description, (ii) time-bounded execution of the generated reducer to obtain the reduced input, and (iii) patch generation where the reduced input, buggy code, and problem description are jointly passed to the LLM. 

To enable rigorous and leakage-free evaluation, we build \dataset, the first APR benchmark that focuses on long test inputs, and we also provide a Python mirror, \datasetpy, with wrong-answer Python submissions. \dataset contains 200 buggy codes from 20 AtCoder tasks released after the training cut-off dates of the evaluated LLMs.
On \dataset, \tool with Qwen2.5-Plus generates syntactically correct reducers for all the 200 bugs, and 95.0\% of those reducers successfully shrink the failure-inducing input by an average of 89.1\%.
\tool with 4 selected LLMs increases their overall pass@10 by up to 53.8\% compared with prompts that embed the whole test, and the advantage holds on \datasetpy. Beyond AtCoder, we validate on 12 OSS-Fuzz crash instances from five projects. Under Docker-grounded validation with Qwen2.5-Plus, reduced inputs raise the micro-average pass@10 to 41.7\% from 25.0\% over the no-test Baseline, and from 16.7\% over the unreduced Origin Test.
Replacing the reduced input for the original input in ChatRepair~\cite{xia2024automated} raises its pass@10 by 21.3\%, and integrating the same plug-in reducer into CREF increases its pass@10 by 2.6\% relative, confirming that \tool can be plugged into existing APR pipelines for an immediate accuracy boost.

% need a brief introduction of results/findings

The main contributions of our work are as follows:

\begin{itemize}[noitemsep,topsep=0pt,leftmargin=2em]
  \item \textbf{Hands-free APR loop with integrating input reduction.} We design a repair framework \tool that prompts an LLM to generate an input reducer to reduce the failure-inducing input and feeds the reduced input to the LLMs for patch generation.
  \item \textbf{Benchmark.} We release the first APR benchmark, \dataset, which contains 200 bugs across 20 different real-world programming tasks, and a Python mirror LFTBench-Py for cross-language validation.
  \item \textbf{Comprehensive evaluations.}
  On \dataset, \tool produces reducers for all bugs and successfully reduces inputs in 95.0\% of cases with an average size reduction of 89.1\%. Reduced tests raise \textit{pass@10} by up to 53.8\% across four LLMs. Comparisons to ddmin-only and pure LLM reducers show clear advantages in both success rate and cost. On LFTBench-Py, the reduced tests also outperform both the no-test and full-test prompts. On OSS-Fuzz, reduced inputs improve crash preservation and increase pass@10 from Origin Test's 16.7\% to 41.7\% under Docker-grounded validation. We further run ablations on prompt length and evidence composition, finding that shortening the failing input yields the largest gains, while appending only output diffs offers limited benefit.
  \item \textbf{Plug-in integration into existing APR systems.} We add \tool as a drop-in reducer to ChatRepair and CREF without changing their logic. ChatRepair’s pass@10 rises by 21.3\% and CREF’s pass@10 increases by 2.6\% relative, showing that \tool is immediately useful as a portable component.
\end{itemize}

\begin{figure*}[h]
  \centering
  \resizebox{.9\linewidth}{!}{%
    \includegraphics[width=\linewidth]{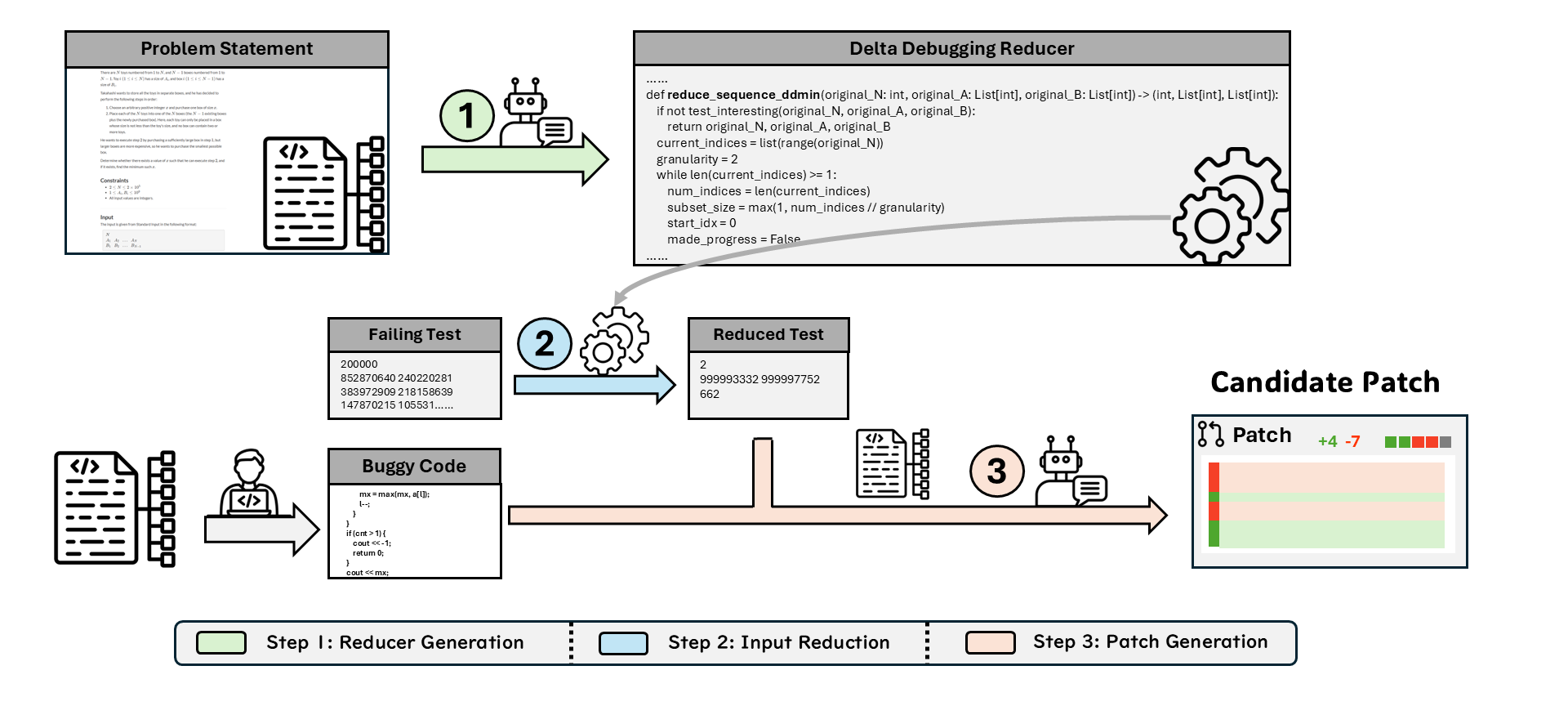}%
  }
  \caption{Overview of \tool.}
  \label{fig:overview}
\end{figure*}

\section{Motivating Example}
\label{sec:motivation}

A single extensive test input can overwhelm an LLM with tokens and conceal the actual defect, which may be located deep within the prompt. This ``lost-in-the-middle'' effect lowers the attention paid to the critical lines and prevents the model from producing an accurate patch. For example, Problem C of AtCoder Beginner Contest 376 makes the issue concrete. The task receives two sorted sequences, $A$ and $B$, and must print the maximum element of $A$ that is not matched by any element of $B$, or $-1$ when more than one element remains unmatched. Figure~\ref{fig:example} sketches the whole scenario.

% Listing~\ref{lst:wa} displays the key part of a wrong-answer submission (No. 65060141). 
The failed submission walks the two sequences from the back, increments \texttt{cnt}, and records \texttt{mx} whenever $a[l]$ is larger than $b[r]$. If $|A|>|B|$, the while-loop exits before the tail of $A$ is checked, so some mismatches slip through.

% \begin{lstlisting}[language=C++,frame=tb,basicstyle=\small,caption={Excerpt of the wrong submission.},label={lst:wa}]
% while (l >= 1 && r >= 1) {
%     if (b[r] >= a[l]) { r--; l--; }
%     else { cnt++; mx = max(mx, a[l]); l--; }
% }
% if (cnt > 1) { cout << -1; return 0; }
% cout << mx;
% \end{lstlisting}

The online judge provides a failure-inducing input that exceeds 3 MB. When Qwen2.5-Coder-7B-Instruct is prompted with the task statement, the buggy code, and this complete file, it adds only a superficial guard and still ignores the trailing elements, so the program continues to fail. The tokens that expose the oversight sit near the midpoint of the 3 MB prompt, far from either end, where the model focuses most of its capacity.

As shown in Figure~\ref{fig:example}, \tool addresses this by inserting an automatic reduction stage before repair. It prompts Qwen2.5-Plus once to write a task-specific Python script that leverages the classical \textit{ddmin} algorithm~\cite{DD} and needs no extra manual effort. Running the LLM-generated reducer reduces the 3 MB test to a three-line counterexample that still forces the buggy and reference programs to diverge.

\iffalse
\begin{lstlisting}[language=Python,frame=tb,basicstyle=\footnotesize,caption={Key parts of the generated reducer.},label={lst:reducer}]
def test_interesting(N, A, B):
    data = encode_case(N, A, B)
    return run("./wa", data) != run("./ac", data)

def reduce_sequence_ddmin(N, A, B):
    current = list(range(N))
    granularity = 2
    # delta debugging loop ...
\end{lstlisting}
\fi

With this compact input, the same repair model introduces a second loop that scans the remaining part of $A$ and updates \texttt{cnt} and \texttt{mx}. The patched program then passes every official test.

Long failure inputs therefore hide the defect and mislead the repair model. Applying a reduction restores focus by keeping only the few tokens that matter, and an LLM can generate the reducer automatically, so the entire process is hands-free. In this example, the reduced prompt improves repair accuracy from an incorrect patch to a fully accepted solution, showing how length control, systematic reduction, and LLM-generated tooling work together to overcome the lost-in-the-middle barrier within APR scenarios.

\section{Approach}\label{sec:approach}

\subsection{Overview}

\tool receives five inputs: the task description $P$, a correct reference solution $A$, a buggy submission $s_w$, the hidden test suite $I$, and one failure-inducing input $i_0$. Its goal is to shrink the failure-inducing input $i_0$ that still distinguishes $A$ from $s_w$, then guide an LLM to repair the bug.

The pipeline proceeds in three stages, shown in Figure~\ref{fig:overview}.
\textbf{Reducer Generation} prompts a code LLM once and returns a customized reducer script that is able to automatically reduce the given failure-inducing input for the task.
\textbf{Input Reduction} executes the generated reducer script under a time limit to shrink the failure-inducing input $i_0$ into a reduced test input $i^{*}$.
\textbf{Patch Generation} embeds $\langle P, s_w, i^{*}\rangle$ in a repair prompt, samples candidate patches, and validates each one against the entire test suite $I$ until a correct program $\hat{s}$ is found or the attempt stops.

Algorithm~\ref{alg:reducefix} lists the full control logic of \tool.

\begin{algorithm}[t]
\caption{\tool\ end-to-end workflow}
\label{alg:reducefix}
\begin{algorithmic}[1]
\REQUIRE the task description $P$, reference correct code $A$, buggy code $s_w$, hidden test suite $I=\{\,i_1,\dots,i_n\,\}$, failure-inducing input $i_0$
\ENSURE Patched program $\hat{s}$ or \textit{failure}
\STATE \textbf{Phase 1: Reducer Generation}
\STATE Build one-shot prompt $\Pi$ using a concrete example
\STATE $R \leftarrow \text{LLM\_Call}(\Pi)$   % $R$ is the generated \texttt{reducer.py} 
\IF{\textsc{StaticCheckFail}$(R)$}
    \STATE \textbf{return} \textit{failure}     % unsafe or malformed reducer
\ENDIF
\STATE \textbf{Phase 2: Input Reduction}
\STATE $i^{\dagger} \leftarrow R.\texttt{reduce}(i_0)$ \COMMENT{within 60 seconds}
\IF{\textsc{Timeout/RE} \textbf{or} $A(i^{\dagger}) = s_w(i^{\dagger})$ \textbf{or} $|i^{\dagger}| \ge |i_0|$}
    \STATE $i^{*} \leftarrow i_0$ \COMMENT{forward the original failing input}
\ELSE
    \STATE $i^{*} \leftarrow i^{\dagger}$ \COMMENT{use the reduced input}
\ENDIF

% \STATE $\rho \leftarrow 1 - \frac{|i^{*}|}{|i_0|}$        % reduction ratio
%
\STATE \textbf{Phase 3: Patch Generation}
\STATE $\hat{s} \leftarrow \text{LLM\_Call}(P,\; s_w,\; i^{*})$
\IF{$\forall\,i\in I : A(i)=\hat{s}(i)$}   % oracle validation over hidden suite
    \STATE \textbf{return} $\hat{s}$
\ENDIF
\STATE \textbf{return} \textit{failure}
\end{algorithmic}
\end{algorithm}

\subsection{Reducer Generation}
\label{subsec:reducer_gen}

Instead of directly reducing test inputs using LLMs, \tool leverages one-shot learning to inject the knowledge of the \emph{ddmin} algorithm~\cite{DD} into the LLM. This enables the model to adapt a well-designed reduction algorithm to various input formats and produce effective, customized reducers.

More specifically, \tool builds one prompt that joins the full task statement $P$, a single-shot example drawn from task \texttt{ABC330D}~\cite{ABC330D} together with its working reducer, and a few I/O pairs from the current task.
The prompt, as shown in Listing~\ref{lst:reducer_prompt}, is sent to Qwen2.5-Plus~\cite{hui2024qwen2} with a temperature of $0$ (greedy decoding). The model returns a reducer $R$, which is a customized reducer derived from the \textit{ddmin} algorithm~\cite{DD}.

\begin{lstlisting}[language=mymarkdown,frame=tb,basicstyle=\small,
                   caption={One-shot reducer prompt fed to Qwen2.5-Plus},
                   label={lst:reducer_prompt}]
First, here is a complete working example for problem {EXAMPLE_PROBLEM_ID_STR}:
# Example Problem Information ({EXAMPLE_PROBLEM_ID_STR})
## Title: {example_problem_title}
## Problem Description (Markdown)
{example_problem_description_md}
## Example `reducer.py` for {EXAMPLE_PROBLEM_ID_STR}
```python
# START --- Example {EXAMPLE_PROBLEM_ID_STR}/reducer.py --- START
{example_reducer_code}
# END --- Example {EXAMPLE_PROBLEM_ID_STR}/reducer.py --- END
```
--- 
Now, using the example above as a reference for structure and helper functions (like `run_program`), please generate the `reducer.py` script for the following target problem: 
# Target Problem Information ({target_problem_id_input})
## Title: {target_problem_title}
## Problem Description (Markdown)
{target_problem_description_md}
Generate the complete `reducer.py` code for problem {target_problem_id_input}. Remember to output only the Python code block.
\end{lstlisting}

% \paragraph{ddmin primer.}
% Let $x_0=i_0$ be the original failure-inducing input. At step $t$, the reducer keeps a candidate $x_t$ and a granularity $g_t\!\ge 2$ (initially $g_0=2$). It partitions $x_t$ into $g_t$ contiguous chunks $B^{(t)}_{1},\dots,B^{(t)}_{g_t}$. The algorithm then evaluates two conditions:

% \begin{align}
% &\exists\,k\in[1,g_t]\;
% \texttt{test\_interesting}\!\bigl(x_t\setminus B^{(t)}_{k}\bigr)=\text{true},
% \label{eq:ddmin-remove}\\[4pt]
% &\exists\,k\in[1,g_t]\;
% \texttt{test\_interesting}\!\bigl(B^{(t)}_{k}\bigr)=\text{true}.
% \label{eq:ddmin-isolate}
% \end{align}

% Equation \eqref{eq:ddmin-remove} asks whether deleting a single chunk still preserves the failure; if so, that chunk is removed and the search restarts with $g_{t+1}=2$. Equation \eqref{eq:ddmin-isolate} checks whether some chunk by itself is enough to trigger the failure; when this holds, the reducer zooms in by letting $x_{t+1}=B^{(t)}_{k}$ and again resets $g_{t+1}=2$. If neither equation is satisfied, the search widens its scope by doubling the granularity, $g_{t+1}=\min(2g_t,|x_t|)$, and repeats. The loop terminates once $g_t>|x_t|$ or both predicates remain false, and the current $x_t$ is returned as the 1-minimal~\footnote{1-minimal refers to that a test input where removing any single part of it makes the failure disappear.} reduced test input $i^{*}$ referenced in Equation~\eqref{eq:minimize}.

\subsection{Input Reduction}

In this stage, the LLM-generated reducer will iteratively shrink the given failure-inducing input $i_0$ while preserving the failure, i.e., the output inconsistency between the accepted reference solution $A$ and the buggy submission $s_w$.
More specifically, the reducer begins by dividing the input into chunks and systematically attempts to produce smaller inputs by removing parts of the original input.
For a produced smaller input $i$, the reducer checks whether it could still cause the inconsistency between the reference correct code $A$ and buggy code $s_w$, i.e., feeding that input $i$ to both $A$ and $s_w$ and evaluating whether their outputs are different. 
If it does, it continues the reduction iteration by replacing the original input with a smaller input; if not, it increases the granularity by splitting the original input into more chunks. This process repeats until no further reduction is possible.

%Reducer will produce a candidate reduced input and check whether a candidate reduced input is still ``interesting'', i.e., the candidate input causes the inconsistency between reference correct code and buggy code, through the predicate
%defined in Equation~\eqref{eq:predicate}.
%A static checker drops any script that tries unsafe actions, such as shell access.
% \begin{equation}
% \texttt{test\_interesting}(i)\;=\;\bigl[A(i)\neq s_w(i)\bigr],
% \label{eq:predicate}
% \end{equation}

%During each iteration, the reducer $R$ produces a candidate input $i$, feeds that input to both $A$ and $s_w$, and evaluates whether their outputs is different. 
%When the two outputs differ, i.e., $\bigl[A(i)\neq s_w(i)\bigr]$, the candidate replaces the current best; otherwise, the change is discarded.
% The reduction loop halts when no further deletion or isolation of any chunk preserves different outputs between $A$ and $s_w$, or when the time limit expires. 

Formally, the reducer seeks  
\begin{equation}
  i^{*}= \arg\min_{i \preceq i_0} |i|
  \quad\text{s.t.}\quad
  A(i)\neq s_w(i),
  \label{eq:minimize}
\end{equation}
where the notation $i \preceq i_0$ means that $i$ is a subsequence of $i_0$, or equivalently,
%$i$ is a subset of $i_0$ in the order of elements.
$i$ can be obtained from $i_0$ by deleting elements in $i_0$ while preserving the relative order of the remaining elements.

The reduction result is a reduced input $i^{*}$. The effectiveness of the reducer is measured by the compression rate, defined as the ratio of the size reduction: 
% $\rho = 1 - \frac{|i^{*}|}{|i_0|}$,
\begin{equation}
  \rho=1-\frac{|i^{*}|}{|i_0|},
  \label{eq:rho}
\end{equation}
where $|i_0|$ and $|i^{*}|$ denote the sizes of the original and reduced input, respectively.

%its quality is measured by the reduction ratio $\rho$ in Equation~\eqref{eq:rho}.
In our settings, the reducer runs reduction iteration for at most 60 seconds.
If the reducer times out, crashes, or cannot shorten the input, the original failure case $i_0$ is forwarded to the next phase.

% Let $m = |i_0|$ be the number of atomic elements (such as lines, bytes, or grid cells) in the origin test. Classical \textsc{ddmin} algorithm~\cite{DD} bounds the worst-case number of predicate evaluations by  
% \begin{equation}
%   T(m) \le m^{2} + 3m ,
%   \label{eq:ddmin-worst}
% \end{equation}
% while the expected time complexity under random failure distributions is $\mathcal{O}(m \log m)$~\cite{DD}.
% With a single test run taking at most $t_{\mathrm{unit}} \!<\! 100$ ms in our environment, even the pessimistic product  
% \begin{equation}
%   t_{\mathrm{unit}} \cdot T(m) < 60 \text{\,s}
%   \label{eq:wall}
% \end{equation}
% remains under the imposed wall-clock ceiling. Empirically, fewer than five percent of reductions hit this safeguard, and the median run consumes less than one percent of the bound in Equation~\eqref{eq:ddmin-worst}.

\subsection{LLM-Guided Repair}
In this stage, \tool first concatenates the task description $P$, the buggy submission $s_w$ and the reduced failure-inducing input $i^{*}$ in a repair prompt, utilizes LLM to samples candidate patches and validates each one until a correct patched program $\hat{s}$ is found.

Because of LLMs’ context length constraints, input prompts have a limited budget.
Due to the fact that reduced test input $i^{*}$ is relatively small, most of them can be inserted into the repair prompt without modification. 
When $i^{*}$ (or its associated output) still exceeds a configurable budget $L$ lines, \tool truncates the literal text shown to the LLM: it keeps the first $\lceil L/2\rceil$ lines and the last $\lfloor L/2\rfloor$ lines, and never splits a line in the middle. Truncation affects only the prompt; the complete file is preserved for compilation and test execution, so semantic correctness is not compromised.

During candidate patch sampling, \tool utilizes the LLM to generate at most $N=10$ candidate patches.
Each candidate must compile within ten seconds and is executed against the full hidden suite $I$. A timeout, runtime error, or wrong answer counts as a failed attempt.  
The repair step succeeds when the first patched program $\hat{s}$ passes the entire test suite $I$: 
\begin{equation}
  \forall\,i\in I,\qquad A(i)=\hat{s}(i).
  \label{eq:repair}
\end{equation}
If no candidate succeeds, the run is reported as a failure.

%Listing~\ref{lst:repair_prompt} shows the exact prompt template. 
If truncation occurred in the inferences of patches, the ellipsis token appears inside the fenced block to signal omitted lines. No other explanatory text is added, keeping the prompt well below typical context limits even on compact LLMs.

% \begin{lstlisting}[language=mymarkdown,frame=tb,basicstyle=\small,
%                    caption={LLM repair prompt used in \tool},
%                    label={lst:repair_prompt}]
% ### Problem Description
% {problem_description}
% ### Your Incorrect Code
% ```cpp
% {wa_code}
% ```
% ### Failing Case
% Input:
% ```
% {reduced_failing_input}
% ```
% Your Output:
% ```
% {wa_output}
% ```
% Expected Output:
% ```
% {expected_output}
% ```
% ### Your Task
% Fix the C++ code to pass ALL test cases (including hidden ones).
% ### Critical Guidelines
% 1. Focus on algorithmic correctness - NO hard-coded values
% 2. Keep complexity reasonable (target $O(N\log N)$ where possible)
% 3. Handle edge cases (empty input, single element, max constraints)
% 4. Use standard C++20 and avoid non-portable extensions
% ### Output Format
% Provide ONLY the complete fixed C++ program inside a single cpp block.
% \end{lstlisting}

\section{Experimental Setup}

\subsection{Models}

% Table \ref{tab:models} summarizes the 4 LLMs evaluated in this work. 

The selection of evaluated LLMs balances two practical factors: the ability to run an LLM locally and the availability of cloud endpoints with competitive token pricing. For the consumer-GPU category, we select \textbf{GLM-4-9B-chat}~\cite{glm2024chatglm} and \textbf{Qwen2.5-Coder-7B-instruct}~\cite{hui2024qwen2}. Both are fully open-source, fit comfortably on a single 24 GB consumer GPU, and therefore incur no API costs. To represent low-cost hosted offerings, we add \textbf{Qwen2.5-Plus}~\cite{hui2024qwen2} and \textbf{DeepSeek-V3}~\cite{liu2024deepseek}. Qwen2.5-Plus is a closed-weight variant of the Qwen2.5 family (the provider does not disclose an exact parameter count) and is priced at \$0.11 per million input tokens and \$0.27 per million output tokens. DeepSeek-V3 is larger (671B parameters, 37B active during decoding) yet still affordable at \$0.27 per million input tokens and \$1.11 per million output tokens. All select LLMs are the standard chat versions rather than the more expensive thinking variants (such as DeepSeek-R1), ensuring a fair and comparable inference budget. This selection enables us to examine how input reduction behaves on both locally deployed LLMs and cost-efficient cloud LLMs.

% \begin{table}[h]
% \centering
% \caption{LLMs used in this study.}
% \footnotesize
% \resizebox{\columnwidth}{!}{
% \begin{tabular}{lccccc}
% \toprule
% \multirow{2.5}{*}{\textbf{Model}} & \multirow{2.5}{*}{\textbf{Params (B)}} & \multirow{2.5}{*}{\textbf{Cutoff}} & \multicolumn{2}{c}{\textbf{\$ Cost per 1M tokens}}\\
% \cmidrule(lr){4-5}
%  &  &  & \textbf{Input} & \textbf{Output}\\
% \midrule
% GLM-4-9B-chat            & 9               & Oct.\,2023 & / & / \\
% Qwen2.5-Coder-7B-instruct         & 7.6             & Mar.\,2024 & / & / \\
% Qwen2.5-Plus             & \textit{N/A}    & Mar.\,2024 & 0.11 & 0.27 \\
% DeepSeek-V3              & 671 & Jun.\,2024 & 0.27 & 1.11 \\
% \bottomrule
% \end{tabular}}
% \label{tab:models}
% \end{table}

\subsection{Benchmark}
\label{subsec:benchmark}

A fair study of input reduction for automated program repair (APR) requires two features that existing datasets lack: (1) \emph{long failure-inducing test inputs} and (2) \emph{low risk of training leakage}. Widely used suites such as Defects4J~\cite{just2014defects4j}, Human-EvalFix~\cite{muennighoffoctopack}, and TutorCode~\cite{yang2024cref}, Codeflaws~\cite{7965296}, and the subset of LeetCode collected by Fan~\cite{fan2023automated} primarily contain only tiny tests, usually a few hundred characters, thus they do not reveal how well an APR pipeline copes when the failing input grows to tens of kilobytes. Meanwhile, most of those benchmarks were released years ago and are drawn from popular open-source projects that large language models have almost certainly seen, which can overstate repair accuracy. To our knowledge, no existing benchmark simultaneously offers long failure-inducing tests and low leakage risk; therefore, we need to build a new benchmark to fill this gap.

To meet the two requirements, the data source must publish every test file, including the largest hidden cases, and it must appear after the training cut-off dates of selected LLMs. LeetCode and Codeforces (including its variant Codeflaws) do not satisfy the first point because they do not share or share only small samples. AtCoder, by contrast, released the full test archives for every Beginner Contest (ABC) up to ABC 377, providing us with large inputs along with an official oracle. Therefore, we select all the satisfied tasks from ABC contests numbered 361 to 377, a span entirely after the knowledge cut-offs of the 4 LLMs we evaluate. From each contest, we retain tasks whose largest official test file is at least 4 KB and whose difficulty level is between C and F, ensuring that the tasks remain solvable for LLMs. To make input diversity explicit, we also categorize the input formats into 6 families and report coverage in Table~\ref{tab:lftbench-input-summary}. For cross-language validation, we additionally prepare \datasetpy, which mirrors the same 20 tasks and reuses the official tests; each task contributes one wrong-answer Python submission. For each task, we manually collect 10 C++ and 1 Python submissions that failed on a large test before July 1, 2025, resulting in 200 and 20 bugs, respectively. The median failing-input size is over 1 MB, and the largest single file exceeds 8 MB, which is large enough to trigger the “lost-in-the-middle” effect.

\begin{table}[t]
  \centering
  \caption{Input-format categories with counts and tasks within \dataset.}
  \label{tab:lftbench-input-summary}
  \resizebox{.9\columnwidth}{!}{
  \begin{tabular}{lcl}
    \toprule
    \textbf{Category} & \textbf{\#Tasks} & \textbf{Tasks} \\
    \midrule
    \multirow{2}{*}{Single sequence} & \multirow{2}{*}{8} & 361C, 367D, 368C, 369D \\
    & & 370D, 373E, 377C, 377F \\
    Multiple sequences & 4 & 374E, 375C, 376C, 376E \\
    Sequence with dependencies & 4 & 364D, 366C, 372E, 371D \\
    2-D matrix & 1 & 363E \\
    Graph & 2 & 362D, 376D \\
    String & 1 & 365D \\
    \bottomrule
  \end{tabular}}
\end{table}

We also include a subset of OSS-Fuzz to cover repository-level defects. OSS-Fuzz continuously exercises project-specific fuzzers across many codebases and yields triaged, reproducible crash inputs with sanitizer stacks, which improves external validity and naturally exposes long and irregular inputs \cite{googleAnnouncingOSSFuzzContinuous}. Although some OSS-Fuzz repositories may overlap with LLM pretraining corpora, we hold the LLM and evaluation instances fixed across strategies, so relative differences between prompt types remain valid despite potential data leakage.

\subsection{Metrics}

A reduction is successful when it completes within 60 seconds, and the resulting input still reproduces the bug. We report (i) the success rate and (ii) the median and average compression rate, as shown in Eq. \ref{eq:rho}.

A repair attempt succeeds when at least one candidate patch passes the full test suite (Eq.~\ref{eq:repair}). For each bug \(b\), we generate \(n_b\) candidates once and observe \(c_b\) passes. We report
\[
\text{pass@}k
=\frac{1}{|\mathcal{B}|}\sum_{b\in\mathcal{B}}
\left(1-\frac{\binom{n_b-c_b}{k_b}}{\binom{n_b}{k_b}}\right),
\quad k_b=\min(k,n_b).
\]
This treats the \(k\) samples as drawn without replacement and ignores ordering. We report \textit{pass@1}, \textit{pass@5}, and \textit{pass@10}.
These three cut-offs align with how developers inspect automated suggestions in practice. Kochhar~\etal~\cite{kochhar2016practitioners} observe that most developers stop using a debugging tool if it does not help within the first five attempts, and Noller~\etal~\cite{noller2022trust} find that few users review more than ten ranked patches. The same thresholds are widely adopted in recent APR work~\cite{huq2022review4repair,fu2022vulrepair,luo2025unlocking,yang2025morepair}. Thus \textit{pass@1} reflects a one-shot setting, while \textit{pass@5} and \textit{pass@10} model realistic batch sizes that can be screened offline without taxing developer patience.

\subsection{Hyperparameters}
\label{hyper-parameters}

Table~\ref{tab:hyperparams} lists every fixed setting used in our experiments. The hyperparameters fall into two operational blocks: reducer generation (including its subsequent \textit{ddmin} search) and repair inference. All values were chosen with small pilot runs on tasks outside the benchmark and kept unchanged throughout the study.

\begin{table}[ht]
\centering
\caption{Key hyper-parameters in \tool.}
\resizebox{\columnwidth}{!}{
\begin{tabular}{@{}llc@{}}
\toprule
\textbf{Stage} & \textbf{Parameter} & \textbf{Setting} \\
\midrule
\multirow{3}{*}{Reducer generation}
& LLM backend & Qwen2.5-Plus \\
& Decoding temperature & 0.0 (greedy) \\
& Wall-clock limit & 60s per reduction \\
\midrule
\multirow{4}{*}{Repair inference}
& \# Samples per bug (pass@k) & $k = \{1,5,10\}$ \\
& Decoding temperature & 0.8 \\
& Compilation timeout & 10s \\
& Execution timeout & 5s per test case \\
\bottomrule
\end{tabular}
}
\label{tab:hyperparams}
\end{table}

To facilitate replication, the full artifact is published at 
\begin{center}
\href{https://github.com/GLEAM-Lab/ReduceFix}{\color{blue}https://github.com/GLEAM-Lab/ReduceFix}
\end{center}

\section{Experiments \& Results}

\subsection{Research Questions}

\begin{itemize}[noitemsep,topsep=0pt,leftmargin=*]
\item \textbf{RQ-1: How reliable are LLM-generated reducers at shrinking failure-inducing inputs?} 
We assess the \tool reduction phase through two metrics: the \emph{success rate} and the \emph{compression ratio}. Both metrics are evaluated against the unreduced original failure-inducing input, a \emph{ddmin}-only approach, and a purely LLM-based input reduction approach.

\item \textbf{RQ-2: Does supplying the reduced counterexample improve LLM-based repair?} 
Using 4 LLMs, we test 3 prompting conditions: Baseline (no failure-inducing input), Origin Test (the full failure-inducing input), and Reduced Test (the reduced input produced by \tool) on \dataset,  and report pass@\textit{k} for $k \in \{1,5,10\}$. To assess cross-language portability, we also run \tool on \datasetpy with Qwen2.5-Plus under the same settings.

\item \textbf{RQ-3: How does prompt composition influence repair accuracy?} 
We study 3 different prompt variants: Reduced Test, Diff Lines, and Reduced + Origin Test, and compare their average lengths and pass@\textit{k}. This analysis distinguishes between the benefits of shorter prompts and the benefits of reduced information.

\item \textbf{RQ-4: Can reduced-input prompting complement existing LLM-based APR pipelines?} 
We integrate the \tool into the ChatRepair~\cite{xia2024automated} and CREF~\cite{yang2024cref} without modifying its logic. Comparing the augmented version against the original one measures whether input reduction provides a drop-in boost for third-party APR systems.

\item \textbf{RQ-5: How well does our approach perform on realistic repository-level repair scenarios?}
We evaluate \tool end-to-end across OSS-Fuzz projects~\cite{googleAnnouncingOSSFuzzContinuous}. For reduction, we compare \tool against the \emph{ddmin}-only and pure-LLM baselines. For repair, we compare pass@\textit{k} across three prompt settings, the same as RQ-2, to assess whether reduced inputs improve repair accuracy.

\end{itemize}

\subsection{RQ-1: Effectiveness of LLM-generated Reducer}
\label{sec:rq1}

\noindent
\textbf{[Objective]:} We evaluate whether an LLM can automatically generate a reducer that preserves program failure while reducing the test input. The goal is to determine both the reliability of the generated reducer and the added value of pairing it with the \textit{ddmin} search, as opposed to relying solely on pure LLM test reduction.

\noindent
\textbf{[Experimental Design]:} We assess the reduction performance on \dataset, which contains 200 buggy C++ submissions drawn from 20 AtCoder Beginner Contest tasks. For each task, we prompt Qwen2.5-Plus once, using a one-shot example (ABC330D) and the problem statement of the task to produce \texttt{reducer.py} (see Section \ref{subsec:reducer_gen}). The script then runs a \textit{ddmin} loop on the full failure-inducing test input to find a smaller input that still causes the output of the buggy and reference programs to differ. To test whether \textit{ddmin} is necessary, we add a baseline named \textbf{pure-LLM}, where the same LLM tries to generate a shorter test input directly.
Furthermore, we introduce a baseline named \textbf{\emph{ddmin}-only}, which directly applies the \emph{ddmin} algorithm to the failure-inducing test input, using spaces and newlines as delimiters.

A reduction is counted as ``Success'' when it finishes on time without errors, and the failure is preserved, and returns a strictly smaller input. %We report the success rate and compression ratio by difficulty and visualize the distribution of compression ratios using a violin plot.

\begin{table}[h]
\centering
\caption{Reducer success and mean compression by difficulty. Compression is averaged over successful reductions.}
\resizebox{\columnwidth}{!}{
\begin{tabular}{l ccc ccc}
\toprule
\multirow{2}{*}{Difficulty (n)} & \multicolumn{3}{c}{Success Rate (\%)} & \multicolumn{3}{c}{Mean/Median Compression Rate (\%)}\\
\cmidrule(lr){2-4}\cmidrule(lr){5-7}
 & \tool & Pure LLM & \textit{ddmin}-only & \tool & Pure LLM & \textit{ddmin}-only \\
\midrule
C (60)        & 100.0 & 63.3 & 30.0 & 84.5/100.0 & 100.0/100.0 & 100.0/100.0 \\
D (80)        & 96.2  & 36.2 & 17.5 & 97.0/100.0 & 100.0/100.0 & 100.0/100.0 \\
E\&F (60)     & 88.3  & 21.7 & 65.0 & 83.0/99.9 & 99.1/99.2  & 84.2/99.9  \\
\midrule
Overall (200) & 95.0  & 40.0 & 35.5 & 89.1/100.0 & 99.8/100.0  & 91.3/100.0  \\
\bottomrule
\end{tabular}
}
\label{tab:rq1-reducer-by-difficulty}
\end{table}
\vspace{-0.5cm}

\begin{table}[h]
\centering
\caption{Reducer success and mean compression by input format. Compression is averaged over successful reductions.}
\resizebox{\columnwidth}{!}{
\begin{tabular}{l ccc ccc}
\toprule
\multirow{2}{*}{Input format (n)} & \multicolumn{3}{c}{Success Rate (\%)} & \multicolumn{3}{c}{Mean/Median Compression Rate (\%)}\\
\cmidrule(lr){2-4}\cmidrule(lr){5-7}
 & \tool & Pure LLM & \textit{ddmin}-only & \tool & Pure LLM & \textit{ddmin}-only \\
\midrule
Single sequence (80)                                                & 97.5 & 36.2 & 52.5 & 85.9/100.0 & 99.9/100.0 & 100.0/100.0 \\
\begin{tabular}{@{}l@{}}Multiple sequences\\(40)\end{tabular} & 100.0 & 50.0 & 22.5 & 76.8/100.0 & 99.5/100.0 & 31.7/0.3  \\
\begin{tabular}{@{}l@{}}Sequence with\\dependencies (40)\end{tabular} & 85.0 & 22.5 & 35.0 & 99.0/100.0 & 100.0/100.0 & 100.0/100.0 \\
2-D matrix (10)                                              & 90.0 & 0.0 & 60.0 & 99.8/99.7 & N/A & 99.8/99.8 \\
Graph (20)                                                   & 95.0 & 65.0 & 0.0  & 100.0/100.0 & 100.0/100.0 & N/A  \\
String (10)                                                  & 100.0 & 90.0 & 0.0  & 100.0/100.0 & 100.0/100.0 & N/A  \\
\midrule
Overall (200)                                                & 95.0 & 40.0 & 35.5 & 89.1/100.0 & 99.8/100.0 & 91.3/100.0 \\
\bottomrule
\end{tabular}
}
\label{tab:rq1-reducer-by-input}
\end{table}
\vspace{-0.5cm}

% \vspace{-0.3cm}
% \begin{table}[h]
% \centering
% \caption{Reducer success rate: pure LLM one-shot vs. \tool\ (LLM + \textit{ddmin}) vs. pure \textit{ddmin}.}
% \resizebox{.8\columnwidth}{!}{
% \begin{tabular}{lcccc}
% \toprule
% \multirow{2.5}{*}{Difficulty} & \multirow{2.5}{*}{Samples} & \multicolumn{3}{c}{Success Rate} \\
% \cmidrule(lr){3-5}
%  &  & Pure LLM & \tool & \textit{ddmin}-only \\
% \midrule
% C    & 60  & 0.63 & 1.00 & 0.30\\
% D  & 80  & 0.36 & 0.96 & 0.18\\
% E\&F    & 60  & 0.22 & 0.88 & 0.65\\
% \midrule
% Overall & 200 & 0.40 & 0.95 & 0.36\\
% \bottomrule
% \end{tabular}
% }
% \label{tab:rq1-reducer-comparison}
% \end{table}
% \vspace{-0.7cm}
% \begin{table}[h]
%  \centering
%  \caption{Compression-rate statistics of \tool.}
%  \resizebox{.65\columnwidth}{!}{
%  \begin{tabular}{lccc}
%   \toprule
%   Difficulty & Samples & Mean (\%) & Median (\%) \\
%   \midrule
%   C  & 60 & 84.5 & 100.0 \\
%   D & 77 & 97.0 & 100.0 \\
%   E\&F  & 53 & 83.0 & 99.9 \\
%   \midrule
%   Overall & 190 & 89.1 & 100.0 \\
%   \bottomrule
%  \end{tabular}
%  }
%  \label{tab:compression_stats}
% \end{table}
% \vspace{-0.7cm}
\begin{table}[h]
\centering
\caption{Token cost comparison on 20 problems.}
\resizebox{.7\columnwidth}{!}{
\begin{tabular}{lccc}
\toprule
\textbf{Method} & \textbf{Pure LLM} & \textbf{\tool} & \textbf{Ratio} \\
\midrule
Input Tokens & 5,687,356 & 94,217 & 0.02 \\
Output Tokens & 1,268 & 20,270 & 15.99 \\
Cost (USD) & \$0.632 & \$0.017 & 0.03 \\
\bottomrule
\end{tabular}
}
\label{tab:token-overall}
\end{table}

\noindent
\textbf{[Experimental Results]:} Using the prompt template in Listing \ref{lst:reducer_prompt}, Qwen2.5-Plus produced a syntactically valid \texttt{reducer.py} for all 200 buggy codes. Hence, all subsequent failures are due to the search phase rather than code-generation errors. Tables~\ref{tab:rq1-reducer-by-difficulty} and \ref{tab:rq1-reducer-by-input} show that the LLM-generated reducers succeed on 95.0\% of the 200 buggy codes. All submissions of C-difficulty tasks are successfully reduced, while D-difficulty tasks and E \& F-difficulty tasks reach 96.2\% and 88.3\% success, respectively. 10 of 200 forwarded the unreduced case to the repair phase, and in all 10 cases, the \emph{ddmin} loop did not yield a shorter candidate before the time limit. The figure in README of our repository lists all 20 synthesized reducers, highlighting their structural decompositions and preserved semantic invariants.

Statistics in Tables \ref{tab:rq1-reducer-by-difficulty} and \ref{tab:rq1-reducer-by-input} report the \emph{compression ratio}, defined as Eq. \ref{eq:rho}, which means the percentage of bytes eliminated by the reducer. Under this definition, larger numbers indicate stronger compression: a value of 100\% indicates that the input was reduced to almost nothing, while 0\% means no reduction at all. The median removal rate remains near 100\% across all difficulty levels, indicating that at least half of the test inputs that induce failures are nearly fully stripped yet still reproduce the bug. Averaged by difficulty, the reducer removes 83.0\% of bytes on E \& F-difficulty tasks, leaving 17.0\% of the original data. Tasks in the C-difficulty and D-difficulty groups achieve even larger average removals of 84.5\% and 97.0\%, respectively.

Tables~\ref{tab:rq1-reducer-by-difficulty} and \ref{tab:rq1-reducer-by-input} show that directly conducting \emph{ddmin} on the input achieves only a 35.5\% success rate. Since input formats are often diverse, directly applying \emph{ddmin} often results in a large number of invalid format attempts and thus exhibits a low reduction success rate. Tables~\ref{tab:rq1-reducer-by-difficulty} and \ref{tab:rq1-reducer-by-input} also show what happens when the LLM is leveraged to generate a shorter test input in a one-shot setting, without the pipeline of \tool. This pure LLM baseline succeeds on only 40.0\% of the bugs overall and 36.2\% of the D-difficulty tasks, and by input type \tool remains robust on graph and string cases where \emph{ddmin}-only fails.

Table~\ref{tab:token-overall} compares the input/output token usage of two reduction strategies on \dataset. The pure LLM approach, where the LLM tries to generate a shorter input in one pass, consumes about 5.7 million input tokens in total. The \tool requires only 0.094 million input tokens for the same dataset. At current API prices, this comparison translates to \$0.632 versus \$0.017, 98\% saving. Two factors drive this saving. First, the reducer script is generated on a per-task basis and then reused for every buggy submission of this task. Second, prompts in \tool do not include the original failure-inducing input, which is often very large in \dataset; they contain only the buggy code and the compact candidate input produced by \textit{ddmin}. The pure LLM baseline must embed the entire long test case in the prompt, which greatly inflates the token cost.

\noindent
\textbf{[Failed Case Study]:} For submission 62869553 of task ABC372E, the student program maintains, for every disjoint-set root, an array that stores the twenty largest vertex identifiers encountered so far. A \texttt{Type 1} query merges two components by copying only the ten largest numbers from the other component, then sorts the twenty numbers in descending order. A \texttt{Type 2} query asks for the $k$-th largest vertex inside the root that contains $v$. When a large component is merged into a smaller one, some very large identifiers disappear; later requests with $k \ge 9$ therefore return incorrect answers.

The generated reducer first applied \textit{ddmin} to discard queries that were not required to trigger the fault, and then renumbered every vertex in the remaining queries to consecutive values $1,2,\dots,|V|$. Because the defect depends on the absolute magnitude of vertex identifiers, this renumbering masked the failure, the interestingness predicate became false, and \textit{ddmin} stopped without shrinking the input.

We can resolve the issue by adding one simple guard. After the reducer renumbers the remaining vertices, it immediately reruns the interestingness test. If the failure no longer reproduces, the script rolls back to the original identifiers for that iteration instead of accepting the renumbered version. This small check, implemented in a few lines of Python, prevents the defect from being masked and allows the reduction process to continue correctly without any other modifications.

\find{{\bf [RQ-1]} \textbf{[Findings]:} (1) \tool successfully reduced on 95\% of the 200 bugs, significantly higher than Pure LLM and \emph{ddmin}-only baselines. (2) Reductions delete 80\%–97\% of inputs on average, leaving a minimal yet still failing input. \textbf{[Insights]:} (1) These results validate the core design of \tool: leveraging LLM to generate a reducer and then driving a systematic \textit{ddmin} search is both essential and efficient. (2) The success of this ``LLM + classic search'' pattern suggests that similar hybrids could improve other code-analysis tasks.}

\subsection{RQ2: Effectiveness of \tool}
\label{sec:rq2}
\noindent\textbf{[Objective]:} We determine whether steering the repair model with the counter-example reduced by \tool improves the accuracy of generating a correct patch compared with (i) no failure-inducing input and (ii) the full failure-inducing input.

\noindent\textbf{[Experimental Design]:} We evaluate whether Reduced Test outperforms both Baseline and Origin Test across pass@\textit{k} for multiple LLMs on \dataset\ (C++), and whether this advantage holds on \datasetpy\ (Python) under the same settings. For every one of the 4 selected LLMs (Qwen2.5-Coder-7B-instruct, GLM-4-9B-chat, Qwen2.5-Plus, DeepSeek-V3), we test three prompting modes:
Baseline (no test), Origin Test(full failure-inducing test), and Reduced Test. All other hyperparameters are fixed (temperature 0.8,  $k \in \{1,5,10\}$, 10 candidate patches per bug), as shown in Table~\ref{tab:hyperparams}. Results are grouped by task difficulty. We also evaluate an end-to-end ddmin-only baseline that uses the reduced input when ddmin succeeds and otherwise falls back to the original failing test, with all other settings unchanged. In addition, for cross-language validation, we evaluate Qwen2.5-Plus on \datasetpy under the same protocol; the reducer succeeds on 18 of 20 Python cases (90\%).

\vspace{-0.3cm}
\begin{table}[h]
\centering
\footnotesize
\caption{Pass@\textit{K} (\%) across 4 LLMs.}
\label{tab:three_strategies}
\resizebox{\columnwidth}{!}{
\begin{tabular}{l ccc ccc ccc}
\toprule
\multirow{2.5}{*}{Difficulty} &
\multicolumn{3}{c}{Baseline (No Test)} &
\multicolumn{3}{c}{Origin Test} &
\multicolumn{3}{c}{Reduced Test}\\
\cmidrule(lr){2-4}\cmidrule(lr){5-7}\cmidrule(lr){8-10}
 & @1 & @5 & @10 & @1 & @5 & @10 & @1 & @5 & @10\\
\midrule
\multicolumn{10}{c}{\textbf{Qwen2.5-Coder-7B-instruct}}\\
C   & 5.5 & 17.1 & 23.3
     & 3.7 & 13.2 & 20.0
     & \textbf{5.5} & 16.7 & \textbf{25.0}\\
D  & 6.4 & 17.4 & 25.0
     & 6.1 & 17.4 & 23.8
     & \textbf{9.9} & \textbf{26.0} & \textbf{36.2}\\
E\&F   & 1.3 & 5.9 & 10.0
     & 1.8 & 7.1 & \textbf{11.7}
     & \textbf{2.3} & \textbf{8.5} & \textbf{11.7}\\
Overall & 4.6 & 13.9 & 20.0
     & 4.1 & 13.1 & 19.0
     & \textbf{6.3} & \textbf{17.9} & \textbf{25.5}\\
\midrule
\multicolumn{10}{c}{\textbf{GLM-4-9B-chat}}\\
C   & 2.8 & 5.8 & 8.3
     & 1.3 & 3.8 & 5.0
     & 2.2 & 4.1 & 5.0\\
D  & 3.5 & 10.6 & 13.8
     & 2.0 & 7.3 & 11.2
     & \textbf{5.8} & \textbf{14.5} & \textbf{20.0}\\
E\&F   & 0.7 & 1.6 & 1.7
     & 0.5 & 1.5 & 1.7
     & \textbf{1.2} & \textbf{1.7} & \textbf{1.7}\\
Overall & 2.5 & 6.5 & 8.5
     & 1.3 & 4.5 & 6.5
     & \textbf{3.3} & \textbf{7.5} & \textbf{10.0}\\
\midrule
\multicolumn{10}{c}{\textbf{Qwen2.5-Plus}}\\
C   & 37.3 & 60.3 & 68.3
     & 31.8 & 53.5 & 63.3
     & 33.7 & 53.5 & 65.0\\
D  & 40.1 & 55.6 & 60.0
     & 39.9 & \textbf{58.7} & 61.3
     & \textbf{43.6} & 58.7 & \textbf{62.5}\\
E\&F   & 22.5 & 39.7 & 48.3
     & 24.0 & 45.7 & 51.7
     & \textbf{25.0} & \textbf{46.5} & \textbf{55.0}\\
Overall & 34.0 & 52.2 & 59.0
     & 32.7 & 53.3 & 59.0
     & \textbf{35.1} & \textbf{53.5} & \textbf{61.0}\\
\midrule
\multicolumn{10}{c}{\textbf{DeepSeek-V3}}\\
C   & 56.5 & 76.3 & 80.0
     & \textbf{56.7} & 75.5 & 78.3
     & 56.3 & \textbf{77.6} & \textbf{81.7}\\
D  & 44.5 & 59.1 & 62.5
     & \textbf{48.9} & 59.2 & 60.0
     & 48.1 & \textbf{61.0} & \textbf{63.7}\\
E\&F   & \textbf{35.0} & \textbf{53.2} & \textbf{58.3}
     & 32.0 & 50.0 & 51.7
     & 32.5 & 51.4 & 56.7\\
Overall & 45.2 & 62.5 & 66.5
     & \textbf{46.1} & 61.3 & 63.0
     & 45.9 & \textbf{63.1} & \textbf{67.0}\\
\bottomrule
\end{tabular}}
\end{table}
\vspace{-0.3cm}
\noindent\textbf{[Experimental Results]:} Across all 4 evaluated LLMs, providing the reduced failure-inducing input consistently raises repair accuracy. Table~\ref{tab:three_strategies} shows that the overall pass@10 climbs from 20\% to 25.5\% for Qwen2.5-Coder-7B-instruct, from 8.5\% to 10.0\% for GLM-4-9B-chat, from 59.0\% to 61.0\% for Qwen2.5-Plus, and from 66.5\% to 67.0\% for DeepSeek-V3 when the Reduced Test prompt replaces the Baseline prompt without any other change.

Gains are strongest on D-difficulty tasks. On D-difficulty tasks, the Reduced Test prompt lifts Qwen2.5’s pass@10 by 44.8\% and pushes GLM-4-9B-chat up by 44.9\%. Tasks with difficulties E\&F see Qwen2.5-Plus rise from 51.7\% to 55.0\%, and DeepSeek-V3 recover part of the loss incurred by the origin test input, moving from 51.7\% to 56.7\%.

In contrast, prompting the unreduced test case often hampers performance. GLM-4-9B-chat’s overall pass@10 falls from 8.5\% with no test case to 6.5\% when the complete input is added, while DeepSeek-V3 drops from 66.5\% to 63.0\% under the same switch. These observations confirm that a focused counterexample, as provided by \tool, supplies the LLM with sufficient evidence to pinpoint the defect.

To test whether the improvement is due to randomness, we leverage MWW tests \cite{mann1947test,wilcoxon1992individual} to further confirm the improvement based on 4 LLMs, returning a two-sided p-value with $<0.05$; therefore, the null hypothesis of equal success probabilities between \tool and the “Origin Test” is rejected, establishing that the gain is statistically significant.
On LFTBench with Qwen2.5-Coder-7B-instruct, leveraging reduced test produced by ddmin-only attains 5.7\% pass@1, 16.1\% pass@5, and 22.0\% pass@10, compared to 6.3\%, 17.9\%, and 25.5\% for \tool; the gap at pass@10 is also statistically significant under a two-sided MWW test (p < 0.05).

On \datasetpy with Qwen2.5-Plus, Reduced Test lifts overall pass@10 to 75.0\% versus 70.0\% for Baseline and 60.0\% for Origin, with the strongest gains on Medium tasks (Table~\ref{tab:py-three-strategies}).

\vspace{-0.3cm}
\begin{table}[H]
\centering
\caption{\tool with Qwen2.5-Plus on \datasetpy by difficulty.}
\label{tab:py-three-strategies}
\resizebox{.9\columnwidth}{!}{
\begin{tabular}{l ccc ccc ccc}
\toprule
\multirow{2.5}{*}{Difficulty} &
\multicolumn{3}{c}{Baseline (No Test)} &
\multicolumn{3}{c}{Origin Test} &
\multicolumn{3}{c}{Reduced Test}\\
\cmidrule(lr){2-4}\cmidrule(lr){5-7}\cmidrule(lr){8-10}
 & @1 & @5 & @10 & @1 & @5 & @10 & @1 & @5 & @10\\
\midrule
C    & 34.8 & 62.2 & 66.7
        & \textbf{38.9} & 62.2 & 66.7
        & 38.3 & \textbf{64.5} & 66.7\\
D  & 53.9 & 69.1 & 75.0
        & 52.5 & 62.2 & 62.5
        & \textbf{61.3} & \textbf{84.4} & \textbf{87.5}\\
E\&F    & \textbf{35.8} & \textbf{66.7} & \textbf{66.7}
        & 18.3 & 40.3 & 50.0
        & 35.2 & 61.5 & \textbf{66.7}\\
\midrule
Overall & 42.7 & 66.3 & 70.0
        & 38.2 & 55.6 & 60.0
        & \textbf{46.5} & \textbf{71.5} & \textbf{75.0}\\
\bottomrule
\end{tabular}}
\end{table}
\vspace{-0.5cm}

\find{{\bf [RQ-2]} \textbf{[Findings]:} (1) Supplying the full failing test often hurts repair accuracy; in multiple LLMs and difficulty levels, it performs worse than the no-test Baseline. (2) \tool is consistently better than Origin Test across all LLMs and difficulty groups, and it also outperforms the Baseline in every overall comparison. (3) When evaluating on \datasetpy, the improvement of \tool is significant, indicating its cross-language robustness. \textbf{[Insights]:} Trimming a long test while preserving its failure signal boosts APR performance, suggesting that controlled input compression may benefit other long-context tasks that must balance prompt length with information retention.}

\subsection{RQ-3: Influence of Prompt Composition on Repair Performance}
\label{sec:rq3}

\noindent
\textbf{[Objective]:} We investigate the distinct influence of two factors within \tool: (i) length reduction (fewer tokens to keep the bug-relevant text within the model’s attention span) and (ii) information selection (retaining the minimal concrete evidence that still exposes the defect).

\noindent
\textbf{[Experimental Design]:} All settings that are unrelated to the prompt remain identical to RQ-2: the benchmark, the Qwen 2.5 Coder-7B-instruct model, the decoding temperature of 0.8, and the sampling times of ten candidate patches per bug. Five prompt variants cover every combination of the two dimensions under study, and an additional control is included with no test information. The \emph{Baseline} prompt contains only the problem statement and the buggy code. \emph{Diff Lines} stays the same length but appends up to ten mismatched output lines, providing sparse evidence without increasing size. \emph{Reduced Test} includes the full input–output pair of the minimized counterexample; it represents the joint action of length control and full information and is the default in \tool. \emph{Origin Test} swaps the reduced input for the unreduced failure-inducing case, inflating the prompt to about 30.6 KB while holding informational content constant, thereby isolating the cost of extra length. \emph{Reduced + Origin} concatenates the reduced and full tests, introducing redundant information as well as maximum length.

% \begin{lstlisting}[language=mymarkdown,frame=tb,basicstyle=\small,caption={Template for \textit{Diff Lines}},label={lst:diff-lines}]
% ### Problem Description
% {full problem text}
% ### Your Incorrect Code
% ```cpp
% {buggy code here}
% ````
% ### Error Summary (diff only)
% Line 1: Got '42', Expected '43'
% Line 2: Got '...', Expected '...'
% ...
% ### Your Task
% Fix the code so that the diff disappears on all tests.
% Return only the complete corrected C++ program in a ```cpp block.
% \end{lstlisting}

For every bug, the LLM is called exactly once with each variant. After generation, we compile and run the candidate patches against the full hidden test suite and record pass@\textit{k} for $k \in \{1,5,10\}$. Mean prompt length is reported alongside pass@\textit{k} for each difficulty group (C, D, E\&F).
\vspace{-0.2cm}
\begin{table}[h]
 \caption{Prompt length statistics for prompting strategies.}
 \centering
 \resizebox{.7\columnwidth}{!}{
 \begin{tabular}{lcc}
  \toprule
  Strategy & Mean Lines & Mean KBs \\
  \midrule
  Baseline  & 130.1     & 3.2 \\
  Origin Test  & 2381.4    & 30.5 \\
  Diff Lines   & 132.6  & 3.2 \\
  Reduced Test  & 514.0   & 6.6 \\
  Reduced and Origin Test & 2637.7 & 36.4 \\
  \bottomrule
 \end{tabular}
 }
 \label{tab:prompt_len}
\end{table}
\vspace{-0.2cm}

\noindent
\textbf{[Experimental Results]:} Table~\ref{tab:prompt_len} confirms that the five prompt variants span more than an order of magnitude in length, from roughly 3 KB for \emph{Baseline} and \emph{Diff Lines} to 36 KB for \emph{Reduced + Origin}. Repair accuracy tracks these length and content differences in a highly systematic way (Table \ref{tab:ablation}). Across the full 200-bug benchmark, \emph{Reduced Test} delivers the best outcomes, reaching 25.5\% pass@10. This score is 5.5 percentage points higher than the \emph{Baseline} that omits test evidence and 5.5 points higher than the \emph{Origin Test} that embeds the unreduced input.

Performance advantages increase as difficulty rises. On the D-difficulty tasks, \emph{Reduced Test} attains 36.2\% pass@10, outperforming both \emph{Diff Lines} (25.0\%) and \emph{Origin Test} (23.8\%). On the hardest E\&F-difficulty, it records 11.7\% pass@10, almost doubling the 6.7\% achieved by \emph{Diff Lines} and more than tripling the 3.3\% scored by \emph{Reduced + Origin}.

Two clear patterns emerge. First, inserting the entire failure-inducing input without reduction inflates the prompt by roughly 30 KB and consistently depresses success, confirming that excessive length dilutes attention. Second, supplying only a terse diff helps little once problems become non-trivial because the LLM lacks a concrete input–output correspondence. The conjunction of compact length and complete counter-example information is therefore essential; either ingredient alone is insufficient. These findings demonstrate that \tool's prompt strategy offers the best trade-off between information richness and context length, a conclusion likely to generalize to other long-context code-related tasks.

\find{{\bf [RQ-3] }\textbf{[Findings]:} (1) Both Diff Lines and Reduced + Origin fall short of the Reduced Test: overall pass@10 drops from 25.5\% to 20.0\% for Diff Lines, and to 19.0\% for Reduced + Origin. (2) Diff Lines is consistently superior to Reduced + Origin; the gap is most pronounced on C-difficulty tasks, where pass@10 reaches 26.7\% versus 23.3\%. \textbf{[Insights]:} Simply shortening the prompt can yield noticeable gains, yet the choice of \emph{how} to compress matters. Experimental results underscore that \tool’s reducer design, which is not length reduction alone, is crucial for the observed accuracy improvements.}

\begin{table}[h]
 \caption{Pass@k (\%) by difficulty under three prompt strategies.}
 \centering
 \resizebox{\columnwidth}{!}{%
 \begin{tabular}{lccccccccc}
  \toprule
  \multirow{2.5}{*}{Difficulty} & \multicolumn{3}{c}{Reduced Test} & \multicolumn{3}{c}{Diff Lines} & \multicolumn{3}{c}{Reduced + Origin}\\
  \cmidrule(lr){2-4}\cmidrule(lr){5-7}\cmidrule(lr){8-10}
   & @1 & @5 & @10 & @1 & @5 & @10 & @1 & @5 & @10 \\
  \midrule
  C  & 5.5 & 16.7 & 25.0 & \textbf{6.7} & \textbf{20.7} & \textbf{26.7} & 4.7 & 16.7 & 23.3 \\
  D & \textbf{9.9} & \textbf{26.0} & \textbf{36.2} & 6.6 & 17.7 & 25.0 & 6.2 & 19.1 & 27.5 \\
  E\&F  & \textbf{2.3} & \textbf{8.5} & \textbf{11.7} & 1.5 & 5.1 & 6.7 & 0.5 & 2.1 & 3.3 \\
  \midrule
  Overall& \textbf{6.3} & \textbf{17.9} & \textbf{25.5} & 5.1 & 14.8 & 20.0 & 4.0 & 13.3 & 19.0 \\
  \bottomrule
 \end{tabular}}
 \label{tab:ablation}
\end{table}

\subsection{RQ-4: Extending \tool to Other APR Pipelines}
\label{sec:rq4}

\noindent
\textbf{[Objective]:} We validate the extensibility of \tool by adding it as a plug-in to the state-of-the-art APR system ChatRepair and CREF, and observing whether the straightforward replacement of the full failure-inducing input with the reduced counter-example leads to a higher patch accuracy.

\noindent
\textbf{[Experimental Design]:} We select ChatRepair~\cite{xia2024automated} and CREF~\cite{yang2024cref} for their open harnesses and drop-in reducer interfaces, which let us replace only the failing test while keeping all other logic unchanged. We fix the sampling budget (k) across settings to isolate the effect of reduced vs. unreduced tests. We keep $k = 10$ across settings to equalize token budgets; increasing k raises absolute pass@k but does not affect our budget-controlled comparisons.

\textsc{ChatRepair}~\cite{xia2024automated} is a conversational repair framework that alternates between a user proxy and an LLM. The first user turn delivers the task description, the buggy C++ program, and a single failing test case. The assistant then proposes a patch, which the testing harness compiles and executes; its verdict (pass or fail) becomes the next user message. A run stops when a patch passes all tests or when the retry budget is exhausted. The original implementation exposes two key hyperparameters: \texttt{MAX\_RETRY}, which limits the number of feedback rounds, and \texttt{length}, which decides how many previous turns are retained in the next prompt.

We evaluate ChatRepair and CREF on the built \dataset benchmark under two prompting modes:

\begin{enumerate}[label=(\alph*),leftmargin=20pt]
 \item \textbf{Origin Test}: Prompting with the full original failure-inducing input test.
 \item \textbf{Reduced Test}: Prompting with the reduced test input produced by the generated reducer.
\end{enumerate}
To isolate the effect of the initial prompt, we fix \texttt{MAX\_RETRY}$=1$ (one feedback round) and set the conversation window size to \texttt{length}$=2$ turns. The LLM samples up to ten candidate patches per bug; we record \textit{pass@k} for $k\!\in\!\{1,5,10\}$ and results are grouped by difficulty.

CREF~\cite{yang2024cref} is a conversational repair framework for programming tutors that iteratively refines a patch using tutor guidance and test feedback across turns. Since CREF expects a tutor-provided hint in its first turn but \dataset does not provide such hints, we omit that turn and run a two-turn variant: we first supply the official AtCoder editorial as the high-level solution description, then provide the failure-inducing test case returned by the harness after validating the first-turn patch as the second-turn feedback.

\begin{table}[h]
 \caption{Repair success of ChatRepair with full versus reduced failure-inducing inputs on \dataset.}
 \centering
 \resizebox{\columnwidth}{!}{
 \begin{tabular}{lcccccc}
  \toprule
  \multirow{2.5}{*}{Difficulty} & \multicolumn{3}{c}{ChatRepair} & \multicolumn{3}{c}{ChatRepair + \tool}\\
  \cmidrule(lr){2-4}\cmidrule(lr){5-7}
   & pass@1 & pass@5 & pass@10 & pass@1 & pass@5 & pass@10\\
  \midrule
  C  & 12.2 & 30.9 & 41.7 & \textbf{14.5} & \textbf{36.6} & \textbf{45.0}\\
  D & 12.1 & 28.2 & 37.5 & \textbf{16.2} & \textbf{35.6} & \textbf{46.2}\\
  E\&F  & 2.3 & 6.8 & 10.0 & \textbf{4.0} & \textbf{12.8} & \textbf{16.7}\\
  \midrule
  Overall & 9.2 & 22.6 & 30.5 & \textbf{12.1} & \textbf{29.0} & \textbf{37.0}\\
  \bottomrule
 \end{tabular}
 }
 \label{tab:rq4-chatrepair}
\end{table}
\begin{table}[h]
 \caption{Repair success of CREF with full versus reduced failure-inducing inputs on \dataset.}
 \centering
 \resizebox{\columnwidth}{!}{
 \begin{tabular}{lcccccc}
  \toprule
  \multirow{2.5}{*}{Difficulty} & \multicolumn{3}{c}{CREF} & \multicolumn{3}{c}{CREF + \tool}\\
  \cmidrule(lr){2-4}\cmidrule(lr){5-7}
   & pass@1 & pass@5 & pass@10 & pass@1 & pass@5 & pass@10\\
  \midrule
  C  & 15.8 & 37.3 & 45.0 & \textbf{17.3} & \textbf{39.5} & \textbf{46.7}\\
  D & 18.8 & 39.4 & 47.5 & \textbf{19.4} & \textbf{39.7} & \textbf{47.5}\\
  E\&F  & 5.5 & 15.6 & 21.7 & \textbf{5.7} & \textbf{16.4} & \textbf{23.3}\\
  \midrule
  Overall & 13.9 & 31.6 & 39.0 & \textbf{14.7} & \textbf{32.7} & \textbf{40.0}\\
  \bottomrule
 \end{tabular}
 }
 \label{tab:rq4-cref}
\end{table}

\noindent
\textbf{[Experimental Results]:} Table \ref{tab:rq4-chatrepair} indicates that replacing the original failure-inducing input with the \tool counter-example improves ChatRepair across every difficulty level. The overall pass@10 rises from 30.5\% to 37.0\%, an absolute gain of 6.5 percentage points and a relative gain of 21.3\%. Looking by difficulty, C-difficulty tasks improve from 41.7\% to 45.0\% (7.9\% relative improvement), D-difficulty tasks from 37.5\% to 46.2\% (23.2\% relative improvement), and the hardest E\&F tasks from 10.0\% to 16.7\% (67.0\% relative improvement). Similar lifts appear at pass@5 and pass@1, confirming that the benefit is robust across different sampling $k$. Likewise, CREF benefits from \tool: overall pass@10 rises from 39.0\% to 40.0\%, with pass@5 from 31.6\% to 32.7\% and pass@1 from 13.9\% to 14.7\% on \dataset\ (Table~\ref{tab:rq4-cref}). These results demonstrate that reducing the supplied failure-inducing test yields a measurable improvement without requiring any other adjustments to existing APR systems.

\find{{\bf [RQ-4]} \textbf{[Findings]:} Integrating \tool with ChatRepair increases the overall pass@10 by 21.3\% relative and with CREF by 2.6\% to the original configuration. \textbf{[Insights]:} \tool can be adopted as a lightweight add-on for any other APR system that leverages test cases, providing an immediate boost in repair effectiveness while requiring no changes to the existing APR pipeline.}

\subsection{RQ-5: Experiments on OSS-Fuzz}
\label{sec:rq5}

\noindent
\textbf{[Objective]:} We aim to evaluate the effectiveness of \tool in realistic software-level repair scenarios beyond C++ programming-exercise benchmarks. Specifically, we test whether \tool can serve as a general, language-agnostic enhancement to LLM-based automated program repair by validating it on real OSS-Fuzz crash instances, assessing both the robustness of the proposed input-reduction mechanism in large, noisy industrial environments and its cross-language applicability across C, C++, and Bash buggy codes.

\noindent
\textbf{[Experimental Design]:} We evaluate a guided repair approach for automated fixing of instances in OSS-Fuzz projects~\cite{googleAnnouncingOSSFuzzContinuous}. We reuse the data and scripts prepared by ARVO~\cite{mei2024arvo} to extract ground-truth patches, failure-inducing inputs, codebases, and issue descriptions, and to evaluate generated patches using pre-built Docker images. The repair pipeline operates iteratively according to the settings in Table~\ref{tab:hyperparams}, except that the reduction phase uses a 600-second time limit to account for Docker test execution overhead. Each LLM invocation receives structured contextual information, including crash diagnostics, complete source files extracted from vulnerable repositories, and failure-inducing input in hexadecimal format. The prompt enforces strict output formatting, including SEARCH/REPLACE blocks with file paths and line numbers for automated patch generation, consistent with prior repository-level repair frameworks~\cite{yang2025enhancing,xia2024agentless}. Each generated patch is immediately validated in the original Docker environment provided by ARVO.
%The ddmin-only uses a two-stage tokenization policy: newline or whitespace tokens for line-oriented inputs, and fixed-size 256-byte chunks with a byte-level fallback for binary or irregular inputs, and we report the best 1-minimal result found within the limit.
For \emph{ddmin}-only, we use whitespace/newline tokens for line-oriented textual inputs as in Section \ref{sec:rq1}; for binary inputs, we adopt fixed 256-byte blocks with a byte-level fallback. We report the best 1-minimal candidate found within the time budget.

To ensure comparability and avoid post-hoc cherry picking, we first filter instances from ARVO whose ground-truth patch modifies exactly one file and whose failure-inducing input exceeds 1 KB. From the pool that satisfies these constraints across five projects (FFmpeg, ImageMagick, Poppler, php-src, MuPDF), we then select the 12 instances with the \textbf{smallest ground-truth patch size} measured by unified diff lines. We compare the effectiveness of reduction across three strategies (\tool, \emph{ddmin}-only, and LLM-generated reduction) by reporting the reduction success rate and compression ratio. Finally, we evaluate the repair accuracy of three strategies (\tool, Baseline without test, and Origin Test) using pass@1, pass@5, and pass@10.

\noindent
\textbf{[Experimental Results]:} 
As shown in Table~\ref{tab:rq5-reduction-comparison} and Table~\ref{tab:repair-pass-at-k}, on 12 OSS-Fuzz crashes spanning FFmpeg, Poppler, PHP-src, and MuPDF, \tool preserved the crash during reduction in 91.7\% of cases versus 75.0\% for \emph{ddmin}-only, a +22.3\% relative gain and an absolute +16.7 points. Mean compression rose from 51.8\% to 56.4\%, a +8.9\% relative gain. Meanwhile, \emph{ddmin}-only failed outright on FFmpeg, while \tool both preserved the crash and reduced the input by 35.4\%. On PHP-src, compression improved from 85.6\% to 97.7\% (+14.1\% relative) with crash preservation at 100\% for both. On Poppler, preservation increased from 50.0\% to 75.0\% (+50.0\% relative) and compression from 22.8\% to 25.4\% (+11.4\% relative). On MuPDF, both preserved the crash, and compression was comparable (-0.8\% relative for \tool).
The pure-LLM approach fails to reduce any of the failure-inducing test inputs, mainly because these inputs are long (exceeding 1 KB) and structurally complex (e.g., binary file), which LLMs cannot effectively simplify.

Under Docker-grounded validation with Qwen2.5-Plus, reduced tests improved end-to-end repair accuracy. On the 12-sample micro average, pass@1 rose from 17.5\% to 19.2\% (+9.7\% relative), pass@5 from 25.0\% to 29.2\% (+16.8\% relative), and pass@10 from 25.0\% to 41.7\% (+66.8\% relative). Using the unreduced Origin Test as the baseline, reduced tests reached 19.2\%, 29.2\%, and 41.7\%, which correspond to +108.7\%, +97.3\%, and +149.7\% relative gains. By project, FFmpeg reached 100.0\% pass@10 from 0\%; MuPDF doubled pass@10 from 20.0\% to 40.0\%; PHP-src raised pass@1 from 60.0\% to 100.0\%; Poppler stayed at 25.0\%; ImageMagick remained unsolved.

In FFmpeg, the ddmin reducer, which deletes fixed-size chunks, often violates the media container length and offset invariants and stops reproducing the failure under the same predicate and harness. In contrast, the LLM-generated reducer applies structure-preserving transformations with adaptive chunking to maintain those invariants before driving ddmin, which preserves the crash and still reduces the input.
\vspace{-0.2cm}
\begin{table}[h]
\centering
\caption{Test case reduction: success rate and compression ratio comparison (Qwen2.5-Plus). SR: Successful Rate, CR: Compress Rate.}
\label{tab:rq5-reduction-comparison}
\resizebox{.75\columnwidth}{!}{
\begin{tabular}{lcccccc}
\toprule
\multirow{2.5}{*}{Project (n)} &
\multicolumn{2}{c}{\textit{ddmin}-only} & \multicolumn{2}{c}{\tool} & \multicolumn{2}{c}{Pure LLM} \\
\cmidrule(lr){2-3}\cmidrule(lr){4-5}\cmidrule(lr){6-7}
& SR & CR & SR & CR & SR & CR \\
\midrule
FFmpeg (1)    &  0.0 &  N/A & 100.0 & 35.4 &  0.0 &  N/A \\
ImageMagick (1)& 100.0 &  4.6 & 100.0 &  5.0 &  0.0 &  N/A \\
MuPDF (5)     & 100.0 & 88.1 & 100.0 & 87.4 &  0.0 &  N/A \\
PHP-src (1)   & 100.0 & 85.6 & 100.0 & 97.7 &  0.0 &  N/A \\
Poppler (4)   &  50.0 & 22.8 &  75.0 & 25.4 &  0.0 &  N/A \\
\midrule
Overall (12)  &  75.0 & 51.8 &  91.7 & 56.4 &  0.0 &  N/A \\
\bottomrule
\end{tabular}
}
\end{table}
\vspace{-0.5cm}
\begin{table}[h]
\centering
\caption{Pass@\textit{K} of \tool (Qwen2.5-Plus) on OSS-Fuzz.}
\label{tab:repair-pass-at-k}
\resizebox{\columnwidth}{!}{
\begin{tabular}{lccccccccc}
\toprule
\multirow{2.5}{*}{Project (n)} &
\multicolumn{3}{c}{Baseline} &
\multicolumn{3}{c}{Origin Test} &
\multicolumn{3}{c}{Reduced Test}\\
\cmidrule(lr){2-4}\cmidrule(lr){5-7}\cmidrule(lr){8-10}
& @1 & @5 & @10 & @1 & @5 & @10 & @1 & @5 & @10 \\
\midrule
FFmpeg (1) &  0.0 &  0.0 &  0.0 &  0.0 &  0.0 &  0.0 & 10.0 & 50.0 & 100.0 \\
ImageMagick (1) &  0.0 &  0.0 &  0.0 &  0.0 &  0.0 &  0.0 &  0.0 &  0.0 &  0.0 \\
MuPDF (5) & 10.0 & 19.9 & 20.0 &  4.0 & 15.6 & 20.0 &  4.0 & 20.0 & 40.0 \\
PHP-src (1) & 60.0 & 100.0 & 100.0 &  0.0 &  0.0 &  0.0 & 100.0 & 100.0 & 100.0 \\
Poppler (4) & 25.0 & 25.0 & 25.0 & 22.5 & 25.0 & 25.0 & 25.0 & 25.0 & 25.0 \\
\midrule
Overall (12) & 17.5 & 25.0 & 25.0 &  9.2 & 14.8 & 16.7 & 19.2 & 29.2 & 41.7 \\
\bottomrule
\end{tabular}
}
\end{table}
\vspace{-0.5cm}
\find{{\bf [RQ-5]} \textbf{[Findings]:} \tool is effective on repository-level repair scenarios. On OSS-Fuzz, \tool preserves crashes more reliably than \emph{ddmin}-only while achieving equal or better compression, and those reductions translate into higher pass@k under evaluation.
The largest improvements in pass@k occur when \emph{ddmin}-only fails or compresses poorly, including in FFmpeg and PHP-src.
\textbf{[Insights]:} 
The evaluation demonstrates the effectiveness of \tool in real-world repository-level repair scenarios. By combining semantic-aware LLM reasoning with traditional reduction techniques, \tool reliably preserves failure behavior while producing more concise and interpretable test cases. These reductions substantially improve downstream repair success, confirming that LLM-guided reduction is not merely a preprocessing step but an essential and practical front-end component of automated repair pipelines at the repository level.
}
\section{Threats to Validity}
\label{sec:threats}

\paragraph{\textbf{Internal validity.}}
The inherent stochasticity of LLMs may pose threats to internal validity.
First, reducer generation is made deterministic by decoding with temperature $=0$; each prompt therefore yields identical code. Second, patch generation retains temperature $=0.8$ so that the model explores diverse fixes. Because developers typically review more than one suggestion, we sample \textit{k} candidate patches per bug and evaluate pass@1, pass@5, and pass@10. Summarizing results through these statistics converts raw sampling noise into a controlled, reproducible measure, thereby mitigating the threat that randomness alone drives the observed accuracy.

\paragraph{\textbf{Construct validity.}}
Improper metrics or biased sampling could distort the study's reflection of real repair difficulty. First, the compression ratio of \tool often reaches 100\%, which hides variation among difficulties. We therefore report both the median and the mean and publish the entire distribution to expose dispersion. Second, pass@\textit{k} may overstate success when a patch is tuned to a partial test set. Every candidate is thus checked against the full official AtCoder archive, rather than a hand-picked or regenerated subset. Third, dataset bias might arise if only select problems are chosen. We include every AtCoder Beginner Contest problem that satisfies the test size and difficulty level filters within a defined date window, thereby removing any scope for cherry-picking. Together, these steps align the measurements with practical repair goals and mitigate the threats to construct validity identified above.

\paragraph{\textbf{External validity.}}
One concern is that \tool might benefit only the pipeline evaluated in this study and fail to transfer to other APR frameworks. To investigate this limitation, we inserted the reducer as a plug-in into ChatRepair and CREF while leaving the rest of its conversation logic unchanged. The modified system achieved a higher pass@10 on the same benchmark, indicating that the method can be integrated into existing APR tools with minimal engineering effort, mitigating the threat of limited applicability. Another concern is about cross-language and scenario generalizability. We validated \tool on \datasetpy and on repository-level OSS-Fuzz and observed improved repair performance in both settings.

\section{Related Work}

\subsection{LLM-based Program Repair}
Program Repair focuses on automatically or semi-automatically fixing software bugs. It aims to reduce the cost and effort of manual debugging by generating patches that correct faulty behavior in code. 
LLMs have recently advanced automated program repair, significantly surpassing traditional rule-based or search-based techniques~\cite{yin_thinkrepair_2024,yang2024cref,xia2022less,yang2025enhancing,yang2025surveyllmbasedautomatedprogram, fan2023automated}. 
ChatRepair is the first work that leverages detailed feedback for each and every patch validated for conversational APR~\cite{xia2024automated}. Following it, many LLM-based repair pipelines embed the failing test case in the prompt to supply bug context~\cite{yang2024cref,kong2024contrastrepair,tang2024code}. However, when the failure-inducing test case is long, the LLMs' context window is exceeded and attention diffuses, producing the ``lost-in-the-middle'' effect and lowering patch accuracy~\cite{yang2024cref,tian2023chatgpt}. Reducing the failure-inducing input while preserving the failure is therefore crucial for efficient, focused repair, especially when the goal is to fine-tune compact LLMs on highly informative examples.
%However, there is no existing input test reduction approach for APR pipelines.
However, no existing work studies test input reduction in the context of APR pipelines.
To our knowledge, our work \tool is the first approach that leverages test input reduction techniques into LLM-based APR.

\subsection{Test Input Reduction}
Reducing test inputs that trigger bugs is crucial for efficient debugging. Delta debugging is the most popular approach for this purpose in software engineering.
Zeller and Hildebrandt initially proposed the first delta debugging algorithm, named \emph{ddmin} \cite{ODD,DD}. Following \emph{ddmin}, HDD~\cite{HDD} and Perses~\cite{Perses} utilize the syntactical structure of the test input to further improve the reduction process.
%by applying \emph{ddmin} on the parse tree of the test input.
ProbDD \cite{ProbDD} introduces a probabilistic model to improve \emph{ddmin}, by estimating the probability of each element being kept in the produced result and prioritizing the reduction of those with high probabilities. GReduce~\cite{GReduce} assumes that the given input is generated by a test generator and applies \emph{ddmin} to the generator’s execution trace to reduce the test input.

Besides the above delta debugging and its derived algorithms, many domain-specific approaches have been proposed for test input reduction on various domains. For example, CReduce~\cite{CReduce}, Vulcan~\cite{10.1145/3586049}, and T-Rec~\cite{10.1145/3690631} reduce source code by iteratively applying pre-defined, well-crafted program transformation rules.
LPR~\cite{10.1145/3650212.3652126} and SimpT5~\cite{wang2025towards} also focus on source code reduction, proposing an LLM-driven technique based on semantic-preserving program transformations. 
ddSMT~\cite{niemetz2013ddsmt} is proposed for conducting delta debugging on SMT formulas. Binary Reduction~\cite{DBLP:conf/sigsoft/KalhaugeP19} is proposed to reduce Java bytecode.
Those approaches are designed for specialized scenarios in which the test inputs take specific forms, including source code and SMT formulas, and are thus not applicable to the competition problems and OSS-Fuzz projects used in our study.

To perform reduction on various types of input formats, such as those in our benchmark tasks, requires manually customizing delta debugging algorithms or designing domain-specific approaches, which is time-consuming and requires domain expertise. In contrast to domain-specific approaches, our work is applicable to a broader range of input types by leveraging LLM’s ability to understand specification descriptions.
Our approach prompts an LLM to automatically generate the input reducer, enabling task-agnostic input minimization that feeds reduced tests directly back to the repair LLMs.

\section{Conclusion}
Our study addresses two persistent gaps that exist between test-case reduction and LLM-driven automated program repair. First, we demonstrate that LLMs can generate a task-specific reducer from a single prompt, thereby freeing developers from the need for manual customization to accommodate heterogeneous input formats. Second, we place this reduction step within the LLM-based repair loop so that the trimmed failing case provides a precise, high-signal prompt for patch generation. Experiments on the newly built \dataset benchmark confirm the value of \tool: inputs shrink by up to 100\%, and repair success climbs by up to 53.8\%. Ablation study shows that the benefit of \tool stems from the combination of brevity and complete failure evidence rather than length trimming alone. An extension experiment further shows that replacing the test input in ChatRepair with the reduced one increases its pass@10 by 21.3\%. These improvements demonstrate that integrating input reduction with LLM-based APR materially advances repair effectiveness.
Evaluation on OSS-Fuzz projects further shows that \tool can improve repair accuracy on industrial-scale projects.

Fully automated reduction unlocks practical applications: programming courses can return minimal counterexamples with fixes to accelerate students' learning, and continuous-integration pipelines can store smaller regression tests and pinpoint faults for developers more efficiently.
Because the reduced input is a plug-in component, future work can extend \tool into other APR frameworks and even broadly long-context LLM tasks that profit from concise but information-rich prompts.

\begin{acks}
This work has been partly supported by National Natural Science Foundation (Grant Number 62273292), China; by Central Leading Local Science and Technology Development Project of Hebei Province (Grant Number 246Z0804G), China; by Innovation Capability Improvement Plan Project of Hebei Province (22567626H), China.
\end{acks}

\balance
\bibliographystyle{ACM-Reference-Format}
\bibliography{sample-base}

@article{le2019automated,
  title={Automated program repair},
  author={Le Goues, Claire and Pradel, Michael and Roychoudhury, Abhik},
  journal={Communications of the ACM},
  volume={62},
  number={12},
  pages={56--65},
  year={2019},
  publisher={ACM New York, NY, USA}
}

@article{xia2024agentless,
  title={Agentless: Demystifying llm-based software engineering agents},
  author={Xia, Chunqiu Steven and Deng, Yinlin and Dunn, Soren and Zhang, Lingming},
  journal={arXiv preprint arXiv:2407.01489},
  year={2024}
}

@article{mei2024arvo,
  title={Arvo: Atlas of reproducible vulnerabilities for open source software},
  author={Mei, Xiang and Singaria, Pulkit Singh and Del Castillo, Jordi and Xi, Haoran and Bao, Tiffany and Wang, Ruoyu and Shoshitaishvili, Yan and Doup{\'e}, Adam and Pearce, Hammond and Dolan-Gavitt, Brendan and others},
  journal={arXiv preprint arXiv:2408.02153},
  year={2024}
}

@misc{googleAnnouncingOSSFuzzContinuous,
	title = {Announcing {OSS}-{Fuzz}: {Continuous} {Fuzzing} for {Open} {Source} {Software}},
	shorttitle = {Announcing {OSS}-{Fuzz}},
	url = {https://testing.googleblog.com/2016/12/announcing-oss-fuzz-continuous-fuzzing.html},
	language = {en},
	urldate = {2025-10-28},
	journal = {Google Testing Blog},
	author = {{Google}},
}

@misc{yang2025surveyllmbasedautomatedprogram,
      title={A Survey of LLM-based Automated Program Repair: Taxonomies, Design Paradigms, and Applications}, 
      author={Boyang Yang and Zijian Cai and Fengling Liu and Bach Le and Lingming Zhang and Tegawendé F. Bissyandé and Yang Liu and Haoye Tian},
      year={2025},
      eprint={2506.23749},
      archivePrefix={arXiv},
      primaryClass={cs.SE},
      url={https://arxiv.org/abs/2506.23749}, 
}

@article{chen2025prometheus,
  title={Prometheus: Unified Knowledge Graphs for Issue Resolution in Multilingual Codebases},
  author={Chen, Zimin and Pan, Yue and Lu, Siyu and Xu, Jiayi and Goues, Claire Le and Monperrus, Martin and Ye, He},
  journal={arXiv preprint arXiv:2507.19942},
  year={2025}
}

@article{tian2023chatgpt,
  title={Is ChatGPT the ultimate programming assistant--how far is it?},
  author={Tian, Haoye and Lu, Weiqi and Li, Tsz On and Tang, Xunzhu and Cheung, Shing-Chi and Klein, Jacques and Bissyand{\'e}, Tegawend{\'e} F},
  journal={arXiv preprint arXiv:2304.11938},
  year={2023}
}

@article{mann1947test,
  title={On a test of whether one of two random variables is stochastically larger than the other},
  author={Mann, Henry B and Whitney, Donald R},
  journal={The annals of mathematical statistics},
  pages={50--60},
  year={1947},
  publisher={JSTOR}
}

@incollection{wilcoxon1992individual,
  title={Individual comparisons by ranking methods},
  author={Wilcoxon, Frank},
  booktitle={Breakthroughs in statistics: Methodology and distribution},
  pages={196--202},
  year={1992},
  publisher={Springer}
}

@inproceedings{liu2020efficiency,
  title={On the efficiency of test suite based program repair: A systematic assessment of 16 automated repair systems for java programs},
  author={Liu, Kui and Wang, Shangwen and Koyuncu, Anil and Kim, Kisub and Bissyand{\'e}, Tegawend{\'e} F and Kim, Dongsun and Wu, Peng and Klein, Jacques and Mao, Xiaoguang and Traon, Yves Le},
  booktitle={Proceedings of the ACM/IEEE 42nd International Conference on Software Engineering},
  pages={615--627},
  year={2020}
}

@inproceedings{pham2015hercules,
  title={Hercules: Reproducing crashes in real-world application binaries},
  author={Pham, Van-Thuan and Ng, Wei Boon and Rubinov, Konstantin and Roychoudhury, Abhik},
  booktitle={2015 IEEE/ACM 37th IEEE International Conference on Software Engineering},
  volume={1},
  pages={891--901},
  year={2015},
  organization={IEEE}
}

@article{yang2025enhancing,
  title={Enhancing Repository-Level Software Repair via Repository-Aware Knowledge Graphs},
  author={Yang, Boyang and Tian, Haoye and Ren, Jiadong and Jin, Shunfu and Liu, Yang and Liu, Feng and Le, Bach},
  journal={arXiv preprint arXiv:2503.21710},
  year={2025}
}

@article{luo2025unlocking,
  title={Unlocking LLM Repair Capabilities in Low-Resource Programming Languages Through Cross-Language Translation and Multi-Agent Refinement},
  author={Luo, Wenqiang and Keung, Jacky Wai and Yang, Boyang and Klein, Jacques and Bissyande, Tegawende F and Tian, Haoye and Le, Bach},
  journal={arXiv preprint arXiv:2503.22512},
  year={2025}
}

@article{liu2023lost,
  title={Lost in the middle: How language models use long contexts},
  author={Liu, Nelson F and Lin, Kevin and Hewitt, John and Paranjape, Ashwin and Bevilacqua, Michele and Petroni, Fabio and Liang, Percy},
  journal={arXiv preprint arXiv:2307.03172},
  year={2023}
}

@article{tang2024code,
  title={Code repair with llms gives an exploration-exploitation tradeoff},
  author={Tang, Hao and Hu, Keya and Zhou, Jin and Zhong, Si Cheng and Zheng, Wei-Long and Si, Xujie and Ellis, Kevin},
  journal={Advances in Neural Information Processing Systems},
  volume={37},
  pages={117954--117996},
  year={2024}
}

@article{kong2024contrastrepair,
  title={Contrastrepair: Enhancing conversation-based automated program repair via contrastive test case pairs},
  author={Kong, Jiaolong and Cheng, Mingfei and Xie, Xiaofei and Liu, Shangqing and Du, Xiaoning and Guo, Qi},
  journal={arXiv preprint arXiv:2403.01971},
  year={2024}
}

@inproceedings{niemetz2013ddsmt,
  title={ddSMT: a delta debugger for the SMT-LIB v2 format},
  author={Niemetz, Aina and Biere, Armin},
  booktitle={Proceedings of the 11th International Workshop on Satisfiability Modulo Theories, SMT},
  pages={8--9},
  year={2013}
}

@inproceedings{just2014defects4j,
  title={Defects4J: A database of existing faults to enable controlled testing studies for Java programs},
  author={Just, Ren{\'e} and Jalali, Darioush and Ernst, Michael D},
  booktitle={Proceedings of the 2014 international symposium on software testing and analysis},
  pages={437--440},
  year={2014}
}

@inproceedings{muennighoffoctopack,
  title={OctoPack: Instruction Tuning Code Large Language Models},
  author={Muennighoff, Niklas and Liu, Qian and Zebaze, Armel Randy and Zheng, Qinkai and Hui, Binyuan and Zhuo, Terry Yue and Singh, Swayam and Tang, Xiangru and Von Werra, Leandro and Longpre, Shayne},
  booktitle={The Twelfth International Conference on Learning Representations}
}

@article{yang2025morepair,
  title={MORepair: Teaching LLMs to Repair Code via Multi-Objective Fine-Tuning},
  author={Yang, Boyang and Tian, Haoye and Ren, Jiadong and Zhang, Hongyu and Klein, Jacques and Bissyande, Tegawende and Le Goues, Claire and Jin, Shunfu},
  journal={ACM Transactions on Software Engineering and Methodology},
  year={2025},
  publisher={ACM New York, NY}
}

@inproceedings{fu2022vulrepair,
  title={VulRepair: a T5-based automated software vulnerability repair},
  author={Fu, Michael and Tantithamthavorn, Chakkrit and Le, Trung and Nguyen, Van and Phung, Dinh},
  booktitle={Proceedings of the 30th ACM Joint European Software Engineering Conference and Symposium on the Foundations of Software Engineering},
  pages={935--947},
  year={2022}
}

@article{huq2022review4repair,
  title={Review4Repair: Code review aided automatic program repairing},
  author={Huq, Faria and Hasan, Masum and Haque, Md Mahim Anjum and Mahbub, Sazan and Iqbal, Anindya and Ahmed, Toufique},
  journal={Information and Software Technology},
  volume={143},
  pages={106765},
  year={2022},
  publisher={Elsevier}
}

@inproceedings{noller2022trust,
  title={Trust enhancement issues in program repair},
  author={Noller, Yannic and Shariffdeen, Ridwan and Gao, Xiang and Roychoudhury, Abhik},
  booktitle={Proceedings of the 44th International Conference on Software Engineering},
  pages={2228--2240},
  year={2022}
}

@inproceedings{kochhar2016practitioners,
  title={Practitioners' expectations on automated fault localization},
  author={Kochhar, Pavneet Singh and Xia, Xin and Lo, David and Li, Shanping},
  booktitle={Proceedings of the 25th international symposium on software testing and analysis},
  pages={165--176},
  year={2016}
}

@article{liu2024deepseek,
  title={Deepseek-v3 technical report},
  author={Liu, Aixin and Feng, Bei and Xue, Bing and Wang, Bingxuan and Wu, Bochao and Lu, Chengda and Zhao, Chenggang and Deng, Chengqi and Zhang, Chenyu and Ruan, Chong and others},
  journal={arXiv preprint arXiv:2412.19437},
  year={2024}
}

@article{hui2024qwen2,
  title={Qwen2. 5-coder technical report},
  author={Hui, Binyuan and Yang, Jian and Cui, Zeyu and Yang, Jiaxi and Liu, Dayiheng and Zhang, Lei and Liu, Tianyu and Zhang, Jiajun and Yu, Bowen and Lu, Keming and others},
  journal={arXiv preprint arXiv:2409.12186},
  year={2024}
}

@inproceedings{DBLP:conf/pldi/KalhaugeP21,
  author       = {Christian Gram Kalhauge and
                  Jens Palsberg},
  editor       = {Stephen N. Freund and
                  Eran Yahav},
  title        = {Logical bytecode reduction},
  booktitle    = {{PLDI} '21: 42nd {ACM} {SIGPLAN} International Conference on Programming
                  Language Design and Implementation, Virtual Event, Canada, June 20-25,
                  2021},
  pages        = {1003--1016},
  publisher    = {{ACM}},
  year         = {2021},
  url          = {https://doi.org/10.1145/3453483.3454091},
  doi          = {10.1145/3453483.3454091},
  bibsource    = {dblp computer science bibliography, https://dblp.org}
}

@article{glm2024chatglm,
  title={Chatglm: A family of large language models from glm-130b to glm-4 all tools},
  author={GLM, Team and Zeng, Aohan and Xu, Bin and Wang, Bowen and Zhang, Chenhui and Yin, Da and Zhang, Dan and Rojas, Diego and Feng, Guanyu and Zhao, Hanlin and others},
  journal={arXiv preprint arXiv:2406.12793},
  year={2024}
}

@inproceedings{xia2024automated,
  title={Automated program repair via conversation: Fixing 162 out of 337 bugs for \$0.42 each using ChatGPT},
  author={Xia, Chunqiu Steven and Zhang, Lingming},
  booktitle={Proceedings of the 33rd ACM SIGSOFT International Symposium on Software Testing and Analysis},
  pages={819--831},
  year={2024}
}

@inproceedings{xia2022less,
  title={Less training, more repairing please: revisiting automated program repair via zero-shot learning},
  author={Xia, Chunqiu Steven and Zhang, Lingming},
  booktitle={Proceedings of the 30th ACM Joint European Software Engineering Conference and Symposium on the Foundations of Software Engineering},
  pages={959--971},
  year={2022}
}

@inproceedings{yang2024cref,
  title={Cref: An llm-based conversational software repair framework for programming tutors},
  author={Yang, Boyang and Tian, Haoye and Pian, Weiguo and Yu, Haoran and Wang, Haitao and Klein, Jacques and Bissyand{\'e}, Tegawend{\'e} F and Jin, Shunfu},
  booktitle={Proceedings of the 33rd ACM SIGSOFT International Symposium on Software Testing and Analysis},
  pages={882--894},
  year={2024}
}

@article{xu2024aligning,
  title={Aligning the Objective of LLM-based Program Repair},
  author={Xu, Junjielong and Fu, Ying and Tan, Shin Hwei and He, Pinjia},
  journal={arXiv preprint arXiv:2404.08877},
  year={2024}
}

@article{DD,
  author    = {Andreas Zeller and
               Ralf Hildebrandt},
  title     = {Simplifying and Isolating Failure-Inducing Input},
  journal   = {{IEEE} Trans. Software Eng.},
  volume    = {28},
  number    = {2},
  pages     = {183--200},
  year      = {2002},
  url       = {https://doi.org/10.1109/32.988498},
  doi       = {10.1109/32.988498},
  bibsource = {dblp computer science bibliography, https://dblp.org}
}

@inproceedings{ODD,
  author       = {Andreas Zeller},
  editor       = {Oscar Nierstrasz and
                  Michel Lemoine},
  title        = {Yesterday, My Program Worked. Today, It Does Not. Why?},
  booktitle    = {Software Engineering - ESEC/FSE'99, 7th European Software Engineering
                  Conference, Held Jointly with the 7th {ACM} {SIGSOFT} Symposium on
                  the Foundations of Software Engineering, Toulouse, France, September
                  1999, Proceedings},
  series       = {Lecture Notes in Computer Science},
  volume       = {1687},
  pages        = {253--267},
  publisher    = {Springer},
  year         = {1999},
  url          = {https://doi.org/10.1007/3-540-48166-4\_16},
  doi          = {10.1007/3-540-48166-4\_16},
  bibsource    = {dblp computer science bibliography, https://dblp.org}
}

@inproceedings{ProbDD,
  author       = {Guancheng Wang and
                  Ruobing Shen and
                  Junjie Chen and
                  Yingfei Xiong and
                  Lu Zhang},
  editor       = {Diomidis Spinellis and
                  Georgios Gousios and
                  Marsha Chechik and
                  Massimiliano Di Penta},
  title        = {Probabilistic Delta debugging},
  booktitle    = {{ESEC/FSE} '21: 29th {ACM} Joint European Software Engineering Conference
                  and Symposium on the Foundations of Software Engineering, Athens,
                  Greece, August 23-28, 2021},
  pages        = {881--892},
  publisher    = {{ACM}},
  year         = {2021},
  url          = {https://doi.org/10.1145/3468264.3468625},
  doi          = {10.1145/3468264.3468625},
  bibsource    = {dblp computer science bibliography, https://dblp.org}
}

@inproceedings{HDD,
  author       = {Ghassan Misherghi and
                  Zhendong Su},
  editor       = {Leon J. Osterweil and
                  H. Dieter Rombach and
                  Mary Lou Soffa},
  title        = {{HDD:} hierarchical Delta Debugging},
  booktitle    = {28th International Conference on Software Engineering {(ICSE} 2006),
                  Shanghai, China, May 20-28, 2006},
  pages        = {142--151},
  publisher    = {{ACM}},
  year         = {2006},
  url          = {https://doi.org/10.1145/1134285.1134307},
  doi          = {10.1145/1134285.1134307},
  bibsource    = {dblp computer science bibliography, https://dblp.org}
}

@inproceedings{Perses,
  author       = {Chengnian Sun and
                  Yuanbo Li and
                  Qirun Zhang and
                  Tianxiao Gu and
                  Zhendong Su},
  editor       = {Michel Chaudron and
                  Ivica Crnkovic and
                  Marsha Chechik and
                  Mark Harman},
  title        = {Perses: syntax-guided program reduction},
  booktitle    = {Proceedings of the 40th International Conference on Software Engineering,
                  {ICSE} 2018, Gothenburg, Sweden, May 27 - June 03, 2018},
  pages        = {361--371},
  publisher    = {{ACM}},
  year         = {2018},
  url          = {https://doi.org/10.1145/3180155.3180236},
  doi          = {10.1145/3180155.3180236},
  bibsource    = {dblp computer science bibliography, https://dblp.org}
}

@article{GReduce,
author = {Ren, Luyao and Zhang, Xing and Hua, Ziyue and Jiang, Yanyan and He, Xiao and Xiong, Yingfei and Xie, Tao},
title = {Validity-Preserving Delta Debugging via Generator Trace Reduction},
year = {2025},
issue_date = {March 2025},
publisher = {Association for Computing Machinery},
address = {New York, NY, USA},
volume = {34},
number = {3},
issn = {1049-331X},
url = {https://doi.org/10.1145/3705305},
doi = {10.1145/3705305},
journal = {ACM Trans. Softw. Eng. Methodol.},
month = feb,
articleno = {65},
numpages = {33}
}

@inproceedings{CReduce,
  author    = {John Regehr and
               Yang Chen and
               Pascal Cuoq and
               Eric Eide and
               Chucky Ellison and
               Xuejun Yang},
  editor    = {Jan Vitek and
               Haibo Lin and
               Frank Tip},
  title     = {Test-case reduction for {C} compiler bugs},
  booktitle = {{ACM} {SIGPLAN} Conference on Programming Language Design and Implementation,
               {PLDI} '12, Beijing, China - June 11 - 16, 2012},
  pages     = {335--346},
  publisher = {{ACM}},
  year      = {2012},
  url       = {https://doi.org/10.1145/2254064.2254104},
  doi       = {10.1145/2254064.2254104},
  bibsource = {dblp computer science bibliography, https://dblp.org}
}

@inproceedings{yin_thinkrepair_2024,
    address = {Vienna Austria},
    title = {{ThinkRepair}: {Self}-{Directed} {Automated} {Program} {Repair}},
    isbn = {979-8-4007-0612-7},
    shorttitle = {{ThinkRepair}},
    url = {https://dl.acm.org/doi/10.1145/3650212.3680359},
    doi = {10.1145/3650212.3680359},
    language = {en},
    urldate = {2025-06-11},
    booktitle = {Proceedings of the 33rd {ACM} {SIGSOFT} {International} {Symposium} on {Software} {Testing} and {Analysis}},
    publisher = {ACM},
    author = {Yin, Xin and Ni, Chao and Wang, Shaohua and Li, Zhenhao and Zeng, Limin and Yang, Xiaohu},
    month = sep,
    year = {2024},
    pages = {1274--1286},
}

@inproceedings{jiang_impact_2023,
    title = {Impact of {Code} {Language} {Models} on {Automated} {Program} {Repair}},
    url = {https://ieeexplore.ieee.org/abstract/document/10172517},
    doi = {10.1109/ICSE48619.2023.00125},
    urldate = {2025-06-11},
    booktitle = {2023 {IEEE}/{ACM} 45th {International} {Conference} on {Software} {Engineering} ({ICSE})},
    author = {Jiang, Nan and Liu, Kevin and Lutellier, Thibaud and Tan, Lin},
    month = may,
    year = {2023},
    note = {ISSN: 1558-1225},
    pages = {1430--1442},
}

@inproceedings{DBLP:conf/sigsoft/KalhaugeP19,
  author       = {Christian Gram Kalhauge and
                  Jens Palsberg},
  editor       = {Marlon Dumas and
                  Dietmar Pfahl and
                  Sven Apel and
                  Alessandra Russo},
  title        = {Binary reduction of dependency graphs},
  booktitle    = {Proceedings of the {ACM} Joint Meeting on European Software Engineering
                  Conference and Symposium on the Foundations of Software Engineering,
                  {ESEC/SIGSOFT} {FSE} 2019, Tallinn, Estonia, August 26-30, 2019},
  pages        = {556--566},
  publisher    = {{ACM}},
  year         = {2019},
  url          = {https://doi.org/10.1145/3338906.3338956},
  doi          = {10.1145/3338906.3338956},
  bibsource    = {dblp computer science bibliography, https://dblp.org}
}

@inproceedings{fan2023automated,
  title={Automated repair of programs from large language models},
  author={Fan, Zhiyu and Gao, Xiang and Mirchev, Martin and Roychoudhury, Abhik and Tan, Shin Hwei},
  booktitle={2023 IEEE/ACM 45th International Conference on Software Engineering (ICSE)},
  pages={1469--1481},
  year={2023},
  organization={IEEE}
}

@misc{ABC330D,
 author= {AtCoder Inc.},
 year  = {2025},
 title = {D - Counting Ls},
 url = "https://atcoder.jp/contests/abc330/tasks/abc330_d",
 note = {Accessed: 2025-07-01}
}

@INPROCEEDINGS{7965296,
  author={Shin Hwei Tan and Jooyong Yi and Yulis and Mechtaev, Sergey and Roychoudhury, Abhik},
  booktitle={2017 IEEE/ACM 39th International Conference on Software Engineering Companion (ICSE-C)}, 
  title={Codeflaws: a programming competition benchmark for evaluating automated program repair tools}, 
  year={2017},
  volume={},
  number={},
  pages={180-182},
  doi={10.1109/ICSE-C.2017.76}}

@inproceedings{10.1145/3650212.3652126,
author = {Zhang, Mengxiao and Tian, Yongqiang and Xu, Zhenyang and Dong, Yiwen and Tan, Shin Hwei and Sun, Chengnian},
title = {LPR: Large Language Models-Aided Program Reduction},
year = {2024},
isbn = {9798400706127},
publisher = {Association for Computing Machinery},
address = {New York, NY, USA},
url = {https://doi.org/10.1145/3650212.3652126},
doi = {10.1145/3650212.3652126},
booktitle = {Proceedings of the 33rd ACM SIGSOFT International Symposium on Software Testing and Analysis},
pages = {261–273},
numpages = {13},
location = {Vienna, Austria},
series = {ISSTA 2024}
}

@article{wang2025towards,
  title={Towards Diverse Program Transformations for Program Simplification},
  author={Wang, Haibo and Xing, Zezhong and Sun, Chengnian and Wang, Zheng and Tan, Shin Hwei},
  journal={Proceedings of the ACM on Software Engineering},
  volume={2},
  number={FSE},
  pages={312--334},
  year={2025},
  publisher={ACM New York, NY, USA}
}

@article{10.1145/3586049,
author = {Xu, Zhenyang and Tian, Yongqiang and Zhang, Mengxiao and Zhao, Gaosen and Jiang, Yu and Sun, Chengnian},
title = {Pushing the Limit of 1-Minimality of Language-Agnostic Program Reduction},
year = {2023},
issue_date = {April 2023},
publisher = {Association for Computing Machinery},
address = {New York, NY, USA},
volume = {7},
number = {OOPSLA1},
url = {https://doi.org/10.1145/3586049},
doi = {10.1145/3586049},
journal = {Proc. ACM Program. Lang.},
month = apr,
articleno = {97},
numpages = {29}
}

@article{10.1145/3690631,
author = {Xu, Zhenyang and Tian, Yongqiang and Zhang, Mengxiao and Zhang, Jiarui and Liu, Puzhuo and Jiang, Yu and Sun, Chengnian},
title = {T-Rec: Fine-Grained Language-Agnostic Program Reduction Guided by Lexical Syntax},
year = {2025},
issue_date = {February 2025},
publisher = {Association for Computing Machinery},
address = {New York, NY, USA},
volume = {34},
number = {2},
issn = {1049-331X},
url = {https://doi.org/10.1145/3690631},
doi = {10.1145/3690631},
journal = {ACM Trans. Softw. Eng. Methodol.},
month = jan,
articleno = {34},
numpages = {31}
}

\end{document}